\documentclass{emulateapj}
\usepackage{apjfonts}
\journalinfo{The Astrophysical Journal, 2011, in press}

\shorttitle{THEORETICAL MASS LOSS RATES OF COOL STARS}
\shortauthors{CRANMER AND SAAR}

\begin{document}

\title{Testing a Predictive Theoretical Model for the
Mass Loss Rates of Cool Stars}

\author{Steven R. Cranmer and Steven H. Saar}
\affil{Harvard-Smithsonian Center for Astrophysics,
60 Garden Street, Cambridge, MA 02138}
\email{scranmer@cfa.harvard.edu,ssaar@cfa.harvard.edu}

\begin{abstract}
The basic mechanisms responsible for producing winds from cool,
late-type stars are still largely unknown.
We take inspiration from recent progress in understanding solar
wind acceleration to develop a physically motivated model of the
time-steady mass loss rates of cool main-sequence stars and
evolved giants.
This model follows the energy flux of magnetohydrodynamic
turbulence from a subsurface convection zone to its eventual
dissipation and escape through open magnetic flux tubes.
We show how Alfv\'{e}n waves and turbulence can produce winds in
either a hot corona or a cool extended chromosphere, and we specify
the conditions that determine whether or not coronal heating occurs.
These models do not utilize arbitrary normalization factors,
but instead predict the mass loss rate directly from a star's
fundamental properties.
We take account of stellar magnetic activity by extending standard
age-activity-rotation indicators to include the evolution of the
filling factor of strong photospheric magnetic fields.
We compared the predicted mass loss rates with observed values for
47 stars and found significantly better agreement than was obtained
from the popular scaling laws of Reimers, Schr\"{o}der, and Cuntz.
The algorithm used to compute cool-star mass loss rates is provided
as a self-contained and efficient computer code.
We anticipate that the results from this kind of model can be
incorporated straightforwardly into stellar evolution calculations
and population synthesis techniques.
\end{abstract}


\section{Introduction}
\label{sec:intro}

All stars are believed to possess expanding outer atmospheres known
as stellar winds.
Continual mass loss has a significant impact on the evolution of
the stars themselves, on surrounding planetary systems, and on the
evolution of gas and dust in galaxies
\citep[see reviews by][]{Du86,LC99,Pu08}.
For example, the Sun's own mass loss was probably an important
factor in the early erosion of atmospheres from the inner planets
of our solar system \citep[e.g.,][]{Wo06,Gu07}.
On the opposite end of the distance scale, a better understanding
of the winds from supergiant stars is leading to new ways of
using them as ``standard candles'' to measure the distances to
other galaxies \citep{Ku10}.
By studying the physical mechanisms that drive stellar winds,
as well as their interaction with processes occurring inside the
stars (convection, pulsation, rotation, and magnetic fields), we are
able to make better quantitative predictions about a wide range of
astrophysical environments.

Over the last half-century, there has been a great deal of research
into possible mechanisms for driving stellar winds on the ``cool
side'' of the Hertzsprung-Russell diagram;
i.e., effective temperatures less than about 8000 K
\citep{HA70,HM80,He88,LB91,Mu96,Wi00,HJ07}.
Despite this work, there is still no agreement about the
fundamental mechanisms responsible for producing these winds.
Many studies of stellar evolution use approximate prescriptions
for mass loss that do not depend on a true physical model of how
the outflow is produced \citep{Rm75,Le10}.
Observational validation of models is made difficult because
mass loss rates similar to that of the solar wind ($\dot{M} \sim
10^{-14}$ $M_{\odot}$~yr$^{-1}$) tend to be too low to be
detectable in most observational diagnostics.

Fortunately, there has been a great deal of recent progress toward
identifying and characterizing the processes that produce our own
Sun's wind.
Self-consistent models of turbulence-driven coronal heating
and solar wind acceleration have begun to succeed in reproducing
a wide range of observations without the need for ad~hoc free
parameters \citep[e.g.,][]{Su06,CvB07,Rp08,VV10,BP11,vB11}.
This progress on the solar front provides a fruitful opportunity
to better understand the fundamental physics of coronal heating
and wind acceleration in other kinds of stars.

The goal of this paper is to construct self-consistent physical
models of cool-star wind acceleration.
These models predict stellar mass loss rates without the
need for observationally constrained normalization parameters,
artificial heating functions, or imposed damping lengths for waves.
We aim to describe time-steady mass outflows from main-sequence
stars with solar-type coronae and from giants with cooler
outer atmospheres.
In principle, then, these models cross the well-known dividing line
\citep{LH79} between stars with and without X-ray emission.
However, there are several types of late-type stellar winds that our
models {\em do not} attempt to explain:
(1) Highly evolved supergiants and asymptotic giant branch (AGB) stars
presumably have winds driven by radiation pressure on dust grains
\citep{LB91,Hf11} and/or strong radial pulsations \citep{Wi00}.
(2) T Tauri stars have polar outflows that may be energized by
magnetospheric streams of infalling gas from their accretion disks
\citep[e.g.,][]{Cr08}.
(3) Blue horizontal branch stars may have line-driven stellar winds
similar to those of O, B, and A type stars \citep{VC02}.

The remainder of this paper is organized as follows.
In Section~\ref{sec:waves} we outline the relevant properties of
Alfv\'{e}n waves and magnetohydrodynamic (MHD) turbulence that
we expect to find in cool-star atmospheres.
Section~\ref{sec:mdot} presents derivations of two complementary
models of mass loss for stars with and without hot coronae, and
also describes how we estimate the total mass loss due to both
gas pressure and wave pressure gradients.
Section~\ref{sec:mag} summarizes how we determine the level of
magnetic activity in a star based on its rotation rate and other
fundamental parameters.
We then give the resulting predictions for mass loss rates of
cool stars in Section~\ref{sec:results} and compare the predictions
with existing observational constraints.
Finally, Section~\ref{sec:conc} concludes this paper with a brief
summary of the major results, a discussion of some of the broader
implications of this work, and suggestions for future improvements.

\section{Alfv\'{e}n Waves in Stellar Atmospheres}
\label{sec:waves}

For several decades, MHD fluctuations have been studied as likely
sources of energy and momentum for accelerating winds from cool
stars \citep[see, e.g.,][]{Ho78,HM80,Dc81,WS91,Ai00,FG06,Su07}.
Specifically, the dissipation of MHD turbulence as a potential
source of heating for the solar wind goes back to \citet{Co68}
and \citet{JD69}.
Despite the fact that other sources of heating and acceleration may
exist, we choose to explore how much can be explained
by restricting ourselves to just this one set of processes.
The ideas outlined here will be applied to both the ``hot'' and
``cold'' models for mass loss described in Section \ref{sec:mdot}.

\subsection{Setting the Photospheric Properties}
\label{sec:waves:photo}

We begin with five fundamental parameters that are assumed to
determine (nearly) all of the other relevant properties of a star:
mass $M_{\ast}$, radius $R_{\ast}$, bolometric
luminosity $L_{\ast}$, rotation period $P_{\rm rot}$, and metallicity.
We also assume that the star's iron abundance, expressed logarithmically
with respect to hydrogen, is a good enough proxy for the abundances of
other elements heavier than helium; i.e.,
[Fe/H]~$\approx  \log (Z/Z_{\odot})$.
For spherical stars, the effective temperature $T_{\rm eff}$ and
surface gravity $g$ are calculated straightforwardly from
\begin{equation}
  \sigma T_{\rm eff}^{4} \, = \, \frac{L_{\ast}}{4\pi R_{\ast}^2}
  \,\, , \,\,\,\,
  g \, = \, \frac{G M_{\ast}}{R_{\ast}^2} \,\, ,
  \label{eq:basic}
\end{equation}
where $\sigma$ is the Stefan-Boltzmann constant and $G$ is the
Newtonian gravitation constant.

We need to know the mass density in the stellar photosphere $\rho_{\ast}$
in order to specify the properties of MHD waves at that height.
The density was computed from the criterion that the Rosseland mean
optical depth should have a value of $2/3$ in the photosphere.
We used the AESOPUS opacity database,%
\footnote{\url{http://stev.oapd.inaf.it/cgi-bin/aesopus}}
which is a tabulation of the Rosseland mean opacity $\kappa_{\rm R}$
as a function of temperature, density, and metallicity
\citep[see][]{MA09}.
An approximate expression for the Rosseland optical
depth $\tau_{\rm R}$ in the photosphere,
\begin{equation}
  \tau_{\rm R} \, = \, \kappa_{\rm R} \rho_{\ast} H_{\ast}
  \, = \, 2/3 \,\, ,
\end{equation}
was solved for $\rho_{\ast}$, where $H_{\ast}$ is the photospheric
value of the density scale height (see below).
We used straightforward linear interpolation to locate the relevant
solutions for $\rho_{\ast}$ as a function of $T_{\rm eff}$, $g$,
and [Fe/H].

\begin{figure}
\epsscale{1.01}
\plotone{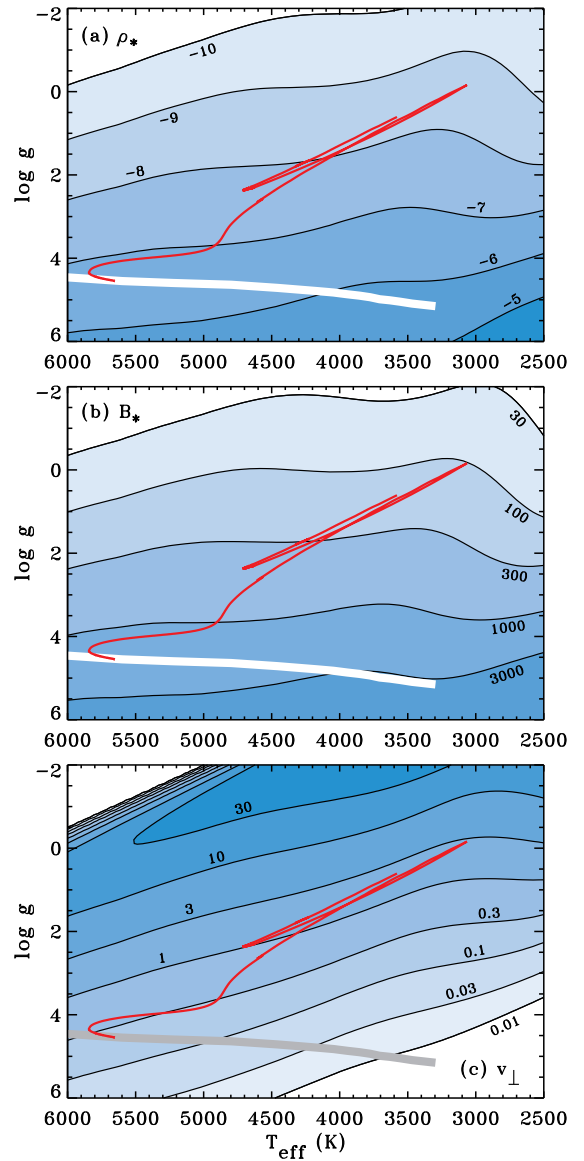}
\caption{Derived photospheric parameters shown as a function of
both $T_{\rm eff}$ and $\log \, g$.
Contour labels denote (a) base-10 logarithm of mass density
$\rho_{\ast}$ in g cm$^{-3}$,
(b) magnetic field strength $B_{\ast}$ in G, and
(c) Alfv\'{e}n wave amplitude $v_{\perp \ast}$ in km s$^{-1}$.
Also shown is a post-main-sequence evolutionary track for
a 1 $M_{\odot}$ star (red curves) and the location of the ZAMS
(white and gray curves).
\label{fig01}}
\end{figure}

Figure \ref{fig01}(a) shows how the photospheric density varies
as a function of $T_{\rm eff}$ and $\log \, g$ under the assumption
of solar metallicity ([Fe/H]~$=0$).
For context we also show the location of the zero-age main sequence
(ZAMS) from the models of \citet{Gi00}, as well as a post-main-sequence
evolutionary track for a 1 $M_{\odot}$ star from the BaSTI%
\footnote{\url{http://albione.oa-teramo.inaf.it/main.php}}
model database \citep{Pi04}.

We used the 2005 release of OPAL plasma equations of state%
\footnote{\url{http://opalopacity.llnl.gov/EOS\_2005/}}
\citep[see also][]{RN02}
to estimate the mean atomic weight $\mu$ in a partially ionized
photosphere.
For the range of parameters appropriate for cool stars,
we found that $\mu$ is primarily sensitive to $T_{\rm eff}$, and
not to gravity or metallicity, so we produced a single parameter fit,
\begin{equation}
  \mu \, \approx \, \frac{7}{4} + \frac{1}{2} \tanh \left(
  \frac{3500 - T_{\rm eff}}{600} \right)
\end{equation}
where $T_{\rm eff}$ is expressed in K.
Other quantities that will be needed later include the photospheric
density scale height, which is given by
\begin{equation}
  H_{\ast} \, = \, \frac{k_{\rm B} T_{\rm eff}}{\mu m_{\rm H} g}
\end{equation}
where $k_{\rm B}$ is Boltzmann's constant and $m_{\rm H}$ is
the mass of a hydrogen atom.
We also need to compute the equipartition magnetic field strength,
\begin{equation}
  B_{\rm eq} \, = \, \sqrt{8\pi P_{\ast}} \, = \,
  \sqrt{\frac{8\pi \rho_{\ast} k_{\rm B} T_{\rm eff}}{\mu m_{\rm H}}}
  \,\, ,
  \label{eq:Beq}
\end{equation}
where $P_{\ast}$ is the photospheric gas pressure.
Because $T_{\rm eff}$ and $\mu$ do not vary over many orders of
magnitude, it is roughly the case that
$B_{\rm eq} \propto \rho_{\ast}^{1/2}$.
However, in all calculations below we compute $B_{\rm eq}$
fully from Equation (\ref{eq:Beq}).
In Section \ref{sec:mag} we describe observations that show the
photospheric magnetic field strength $B_{\ast}$ is roughly linearly
proportional to $B_{\rm eq}$ for many stars.
The measurements determine the constant of proportionality, and we use
\begin{equation}
  B_{\ast} \, = \, 1.13 \, B_{\rm eq}
  \label{eq:Bstar}
\end{equation}
in the remainder of this paper.
Figure \ref{fig01}(b) shows $B_{\ast}$ as a function of
$T_{\rm eff}$ and $\log \, g$.

We consider MHD waves that are driven by turbulent convective motions
in the stellar interior.
The original models of wave generation from turbulence
\citep[e.g.,][]{Li52,Pr52,St67} dealt mainly with acoustic waves
in an unmagnetized medium.
More recently, however, it has been shown that when a stellar
atmosphere is filled with magnetic flux tubes, the dominant carrier
of wave energy should be transverse kink-mode oscillations \citep{Mz02a}.
When the magnetic flux tubes extend above the stellar surface and
expand to fill the volume, the kink-mode waves become shear
Alfv\'{e}n waves \citep[see][]{CvB05}.

\begin{figure}
\epsscale{1.17}
\plotone{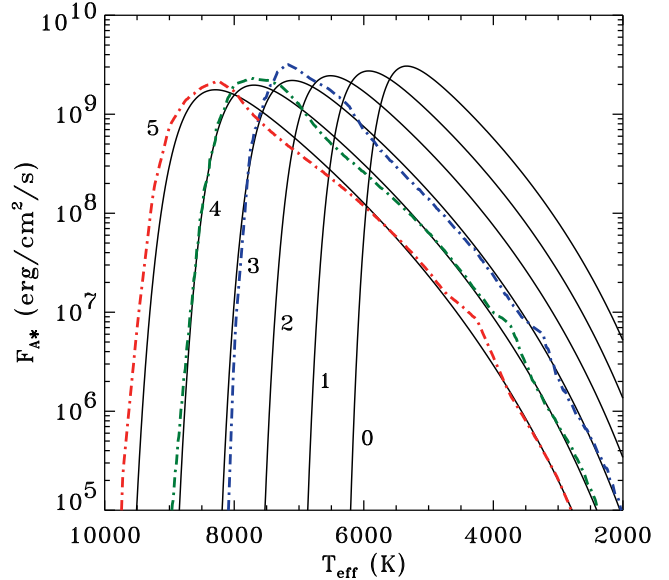}
\caption{Comparison between the \citet{Mz02a} numerical models
(dot-dashed curves) and analytic fits (solid curves) for photospheric
transverse wave energy fluxes $F_{\rm A \ast}$ as a function of
effective temperature and photospheric gravity.
Numerical labels denote $\log \, g$ for each curve.
\label{fig02}}
\end{figure}

We utilize the model results of \citet{Mz02a} to estimate the
flux of energy in kink/Alfv\'{e}n waves in stellar photospheres.
For simplicity, we used only the simulations of \citet{Mz02a} with
their standard parameter choices: a mixing length parameter of
$\alpha = 2$ and a constant magnetic field strength that is 0.85
times the equipartition field strength.
Our analytic fit to the results shown in their Figure 8 is
\begin{equation}
  F_{\rm A \ast} \, = \, F_{0}
  \left( \frac{T_{\rm eff}}{T_0} \right)^{\varepsilon}
  \exp \left[ - \left( \frac{T_{\rm eff}}{T_0} \right)^{25} \right]
  \label{eq:FAfit}
\end{equation}
where the dependence on $\tilde{g} = \log \, g$ is given by
\begin{equation}
  \frac{F_0}{10^{9} \,\,
  \mbox{erg} \,\,\, \mbox{cm}^{-2} \,\,\, \mbox{s}^{-1}}
  \, = \, 5.724 \,\, \exp
  \left( - \frac{\tilde{g}}{11.48} \right) \,\, ,
\end{equation}
\begin{equation}
  \frac{T_0}{1000 \,\, \mbox{K}}
  \, = \, 5.624 + 0.6002 \, \tilde{g}  \,\, ,
\end{equation}
\begin{equation}
  \varepsilon \, = \, 6.774 + 0.5057 \, \tilde{g} \,\, .
\end{equation}
These fits are similar in form to those given by \citet{Fw11} for
longitudinal MHD waves.
Figure \ref{fig02} shows a comparison between the above fitting formula
and the plotted results of \citet{Mz02a} for $\log \, g = 3$, 4, and 5.
The behavior of $F_{\rm A \ast}$ for lower values of $\log \, g$ was
not given by \citet{Mz02a}, but similar results were found for a wider
range of gravities by \citet{Um96} for acoustic waves.

We used the kink-mode energy flux to determine the transverse
velocity amplitude $v_{\perp}$ of Alfv\'{e}n waves in the photosphere.
The flux is defined as
\begin{equation}
  F_{\rm A \ast} \, = \, \rho_{\ast} v_{\perp \ast}^{2}
  V_{{\rm A} \ast}
\end{equation}
with $V_{{\rm A} \ast} = B_{\ast} / (4\pi \rho_{\ast})^{1/2}$ being
the photospheric Alfv\'{e}n speed.
The above expression is not exact for waves undergoing strong
reflection \citep[see, e.g.,][]{HO80}, but it ends up giving a
similar prediction for the height variation of $v_{\perp}$ in the
corona that would come from a more accurate non-WKB model
\citep{CvB05}.
Figure \ref{fig01}(c) shows how $v_{\perp \ast}$ varies as a function
of $T_{\rm eff}$ and $\log \, g$ for solar metallicity stars.

For the well-observed case of the Sun, we know that most of the
photospheric magnetic field is concentrated into small
(100--200 km diameter) flux tubes concentrated in the intergranular
downflow lanes \citep{So93,BT01}.
The field strength in these tubes is close to equipartition,
with $B_{\ast} \approx 1400$ G.
However, these flux tubes have a {\em filling factor} $f_{\ast}$
in the photosphere of about 0.1\% to 1\%, so the spatially
averaged magnetic flux density $B_{\ast} f_{\ast}$ is only of
order 1--10 G \citep{SH89}.

\subsection{Radial Evolution of Waves and Turbulence}
\label{sec:waves:rad}

\begin{figure}
\epsscale{1.045}
\plotone{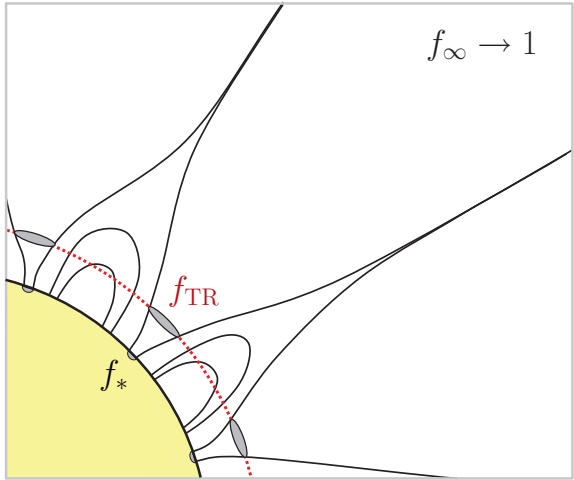}
\caption{Summary illustration of flux tube expansion on a
representative cool star.
The filling factor (for open magnetic flux tubes) grows from $f_{\ast}$
in the photosphere to $f_{\rm TR}$ at the transition region, and to an
asymptotic value $f_{\infty} \rightarrow 1$ at large distances.
The dimensions in this sketch are not drawn to scale.
\label{fig03}}
\end{figure}

Figure \ref{fig03} illustrates the stellar magnetic field geometry
that we assume to exist above the surface of a cool star.
Flux tubes that are open to the stellar wind\footnote{%
The presumed non-existence of magnetic monopoles implies that
``open'' field lines must eventually be closed far from the star,
presumably via interactions with the larger-scale interstellar
field \citep{Dv55}.}
have a cross-sectional area $A(r)$ that expands monotonically with
increasing radial distance $r$ from the star.
The condition $\nabla \cdot {\bf B} = 0$ demands that the product
of $A$ and the magnetic field strength $B$ remains constant.
Thus, $B(r)$ inside a flux tube decreases monotonically, from its
photospheric value of $B_{\ast}$, with increasing distance.
We normalize $A$ such that at a given distance the total stellar
surface area covered by open flux tubes is defined to be
\begin{equation}
  A \, = \, 4 \pi r^{2} f   \,\, .
\end{equation}
The dimensionless filling factor $f$ tends to increase with height to
an asymptotic value of 1 as $r \rightarrow \infty$
\citep[see also][]{Cu99}, but its increase is not necessarily monotonic.
We do not explicitly consider the properties of closed magnetic
``loops'' on the stellar surface, but the radial variation of
$f(r)$ takes into account their presence.

For each star, we intend to specify $f_{\ast}$ on the basis of
either direct measurements or empirical scaling relations.
The model described in Section \ref{sec:mdot:hot} also requires
specifying the value of $f$ at the sharp transition region (TR)
between the cool chromosphere and hot corona.
We generally know that $f_{\ast} < f_{\rm TR} < 1$, so in
the absence of better information we will apply the assumption that
that $f_{\rm TR} = f_{\ast}^{\theta}$, where $\theta$ is a
dimensionless constant between 0 and 1.
For the solar wind models of \citet{CvB07} the exponent $\theta$
ranges between about 0.3 and 0.5.

Alfv\'{e}n waves propagate up from the stellar photosphere, partially
reflect back down toward the Sun, develop into strong MHD turbulence,
and dissipate gradually \citep{Ve91,Mt99,CvB05}.
Temporarily ignoring the reflection and turbulent cascade, the
overall energy balance of an Alfv\'{e}n wave train is governed by
the conservation of wave action.
We define the flux of wave action $\tilde{S}$ as
\begin{equation}
  \tilde{S} \, \equiv \,
  \rho v_{\perp}^{2} V_{\rm A} (1 + M_{\rm A})^{2} A \, = \,
  \mbox{constant}
  \label{eq:action}
\end{equation}
where $M_{\rm A} = u / V_{\rm A}$ is the Alfv\'{e}n Mach number
and $u$ is the radial outflow speed of the wind
\citep[see, e.g.,][]{J77,TM95}.
Close to the stellar surface, where $M_{\rm A} \ll 1$, this condition
is equivalent to energy flux conservation
($F_{\rm A} A = \mbox{constant}$).
In any case, the constant value of $\tilde{S}$ in
Equation (\ref{eq:action}) is known for each star because the
conditions at the photosphere are known (and it is also valid to
assume $M_{\rm A} \rightarrow 0$ there as well).
The behavior of the wave amplitude as a function of density varies
from $v_{\perp} \propto \rho^{-1/4}$ close to the star to
$v_{\perp} \propto \rho^{+1/4}$ at larger distances.

The waves gradually lose energy due to turbulent dissipation, but 
for locations reasonably close to the stellar surface---e.g., the
region shown in Figure \ref{fig03}---it is not a bad approximation to
use the undamped form of wave action conservation to compute the
radial dependence of $v_{\perp}$ \citep[see][]{CvB05}.
Wave damping gives rise to plasma heating, and we adopt a
phenomenological heating rate that is consistent with the total
energy flux that cascades from large to small eddies.
This rate is constrained by the properties of the Alfv\'{e}nic
fluctuations at the largest scales, and it does not specify the exact
kinetic means of dissipation once the energy reaches the smallest scales.
Dimensionally, it is similar to the rate of cascading energy
flux derived by \citet{vK38} for isotropic hydrodynamic turbulence.
The volumetric heating rate is given by
\begin{equation}
  Q \, = \, \frac{\tilde{\alpha} \rho v_{\perp}^3}{\lambda_{\perp}}
  \label{eq:Qbasic}
\end{equation}
\citep{Ho86,Hs95,ZM90,Mt99,Dm02}.
The dimensionless efficiency factor $\tilde{\alpha}$ depends on the
local degree of wave reflection and is discussed further below.
The perpendicular length scale $\lambda_{\perp}$ is an effective
correlation length for the largest eddies in the turbulent cascade.

MHD turbulence occurs only when there exist counter-propagating
Alfv\'{e}n wave packets along a flux tube.
The star naturally creates upward waves, and we assume that linear
reflection gives rise to downward waves \citep{FP58}.
We specify the ratio of downward to upward wave amplitudes by the
effective reflection coefficient ${\cal R}$, and the efficiency
factor $\tilde{\alpha}$ is given by
\begin{equation}
  \tilde{\alpha} \, = \,
  \alpha_{0} \frac{{\cal R} (1 + {\cal R}) \sqrt{2}}
  {(1 + {\cal R}^{2})^{3/2}}
\end{equation}
\citep[see, e.g.,][]{CvB07}.
At the photospheric lower boundary, we assume total reflection with
${\cal R} = 1$ and thus $\tilde{\alpha} = \alpha_0$.
Higher in the stellar atmosphere, we use the low-frequency limiting
expression of \citet{Cr10},
\begin{equation}
  {\cal R} \, \approx \, \frac{V_{\rm A} - u_{\infty}}
  {V_{\rm A} + u_{\infty}} \,\, ,
  \label{eq:RefLow}
\end{equation}
where the wind's terminal speed is $u_{\infty}$ and we also assume
that $M_{\rm A} \ll 1$ in the atmosphere.
This expression also assumes that the wind speed at
the point where $M_{\rm A} = 1$ (presumably far from the stellar
surface) is roughly equal to $u_{\infty}$.
We also set $\alpha_{0} = 0.5$ based on the turbulent transport
models of \citet{Br09}.

\section{Models for Mass Loss}
\label{sec:mdot}

In this section we present two complementary descriptions of cool-star
mass loss that make use of the Alfv\'{e}n wave properties discussed above.
Supersonic winds can be driven by either gas pressure in a hot corona
(Section \ref{sec:mdot:hot}) or wave pressure in a cool, extended
chromosphere (Section \ref{sec:mdot:cold}).
We first investigate each idea by assuming the other one is negligible,
and then we explore how to incorporate both processes together
(Section \ref{sec:mdot:both}).

\subsection{Hot Coronal Mass Loss}
\label{sec:mdot:hot}

If the turbulent heating given by Equation (\ref{eq:Qbasic}) is
sufficient to produce a hot ($T \gtrsim 10^{6}$ K) corona, then
the plasma's high gas pressure gradient may provide enough outward
acceleration to produce a transition from a subsonic (bound)
state near the star to a supersonic (outflowing) state at larger
distances \citep{P58}.
In this section we estimate the mass loss rate $\dot{M}$ of such a
gas-pressure-driven stellar wind.

We begin by computing $Q_{\ast}$ in the photosphere using
$\rho_{\ast}$ and $v_{\perp \ast}$ in Equation (\ref{eq:Qbasic}).
For the Sun, we have the observational constraint that
$\lambda_{\perp \ast}$ must be about the size of the granular motions
that jostle the flux tubes (i.e., roughly 100--1000 km).
For other stars we can assume that the horizontal scale of
granulation remains proportional to the photospheric pressure scale
height \citep{Rf04}.
Thus, we use
\begin{equation}
  \lambda_{\perp \ast} \, = \, \lambda_{\perp \odot} \,
  \frac{H_{\ast}}{H_{\odot}} \,\, ,
  \label{eq:lamstar}
\end{equation}
where $H_{\odot} = 139$~km and the
models of \citet{CvB05} were used to set the solar normalization of
the correlation length to $\lambda_{\perp \odot} = 300$ km.

In the photosphere, we assume the turbulent heating is swamped by
radiative gains and losses that are determined by the conditions of
local thermodynamic equilibrium (LTE), and the temperature is set
by those processes alone.
At larger heights in the flux tube, the turbulent heating
$Q$ begins to have an effect.
We define the {\em chromosphere} as the region in which $Q$ is balanced
by radiative losses.
As one increases in height, however, the density drops to the point
where radiative losses alone can no longer balance the imposed
heating rate; this occurs at the sharp TR between chromosphere and
corona.
(See Section \ref{sec:mdot:cold} for cases where this transition does
not occur at all.)

In the region between the photosphere and the TR, we assume that
the wind flow speed is sufficiently sub-Alfv\'{e}nic 
such that $v_{\perp} \propto \rho^{-1/4}$.
We also assume that $\lambda_{\perp}$ scales with the
transverse size of the magnetic flux tube, so that
$\lambda_{\perp} \propto A^{1/2} \propto B^{-1/2}$ \citep{Ho86}.
Thus, Equation (\ref{eq:Qbasic}) can be rewritten as
\begin{equation}
  \frac{Q_{\rm TR}}{Q_{\ast}} \, = \,
  \frac{\tilde{\alpha}_{\rm TR}}{\tilde{\alpha}_{\ast}}
  \left( \frac{\rho_{\rm TR}}{\rho_{\ast}} \right)^{1/4}
  \left( \frac{B_{\rm TR}}{B_{\ast}} \right)^{1/2}
\end{equation}
where $\tilde{\alpha}_{\ast} = 0.5$ and all other photospheric
quantities are assumed to be known.
We also know that
\begin{equation}
  \frac{B_{\rm TR}}{B_{\ast}} \, = \,
  \frac{f_{\ast}}{f_{\rm TR}} \, \approx \, f_{\ast}^{1-\theta}
\end{equation}
where the last approximation holds if there is a universal
relationship between $f_{\ast}$ and $f_{\rm TR}$ as speculated in
Section \ref{sec:waves:rad} above.

Just below the TR, the heating is just barely balanced by radiative
cooling.
In the optically thin limit, radiative cooling behaves as
$Q_{\rm cool} = - n^{2} \Lambda(T)$, where $n$ is the number density
in the fully ionized TR region.
Let us then assume that $Q_{\rm TR} = \max | Q_{\rm cool} |$, where
\begin{equation}
  \max | Q_{\rm cool} | \, = \,
  \frac{\rho_{\rm TR}^{2} \Lambda_{\rm max}}{m_{\rm H}^2}  \,\, .
  \label{eq:maxcool}
\end{equation}
The quantity $\Lambda_{\rm max}$ is the absolute maximum of the
radiative loss curve $\Lambda (T)$, and it occurs roughly at
$T_{\rm TR} = 2 \times 10^{5}$ K.
The value of $\Lambda_{\rm max}$ depends on metallicity.
To work out its dependence on $Z$, we
computed a number of radiative loss curves for different metal
abundances using version 4.2 of the CHIANTI atomic database
\citep{Yo03} with collisional ionization balance \citep{Mz98}.
We started with a traditional \citep{GS98} solar abundance
mixture ($Z/Z_{\odot} = 1$) and then recomputed $\Lambda(T)$
by varying the metal abundance ratio $Z/Z_{\odot}$ between 0 and 10.
We found that the maxima of the curves were fit well by the
following parameterized function,
\begin{equation}
  \frac{\Lambda_{\rm max}}{10^{-23} \, \mbox{erg} \,\, \mbox{cm}^{3}
  \,\, \mbox{s}^{-1}} \, \approx \, 7.4 +
  42 \left( \frac{Z}{Z_{\odot}} \right)^{1.13}  \,\, .
  \label{eq:Lammax}
\end{equation}
Other examples of the metallicity dependence of $\Lambda(T)$ have been
given by, e.g., \citet{BH89} and \citet{Gn07}.
We have ignored any possible differences between a star's photospheric
metal abundances and those in the low corona, although such differences
have been measured in some cases \citep{Te10}.

With the above assumptions, we solve for the TR density,
\begin{equation}
  \rho_{\rm TR} \, = \, \left[
  \frac{\tilde{\alpha}_{\rm TR} Q_{\ast} m_{\rm H}^{2}}
  {\tilde{\alpha}_{\ast} \rho_{\ast}^{1/4} \Lambda_{\rm max}}
  \right]^{4/7} {f_{\ast}}^{2(1-\theta)/7}
  \label{eq:rhoTR}
\end{equation}
and we also derive the heating rate at the TR to be
\begin{equation}
  Q_{\rm TR} \, = \, \left(
  \frac{\tilde{\alpha}_{\rm TR} Q_{\ast}}{\tilde{\alpha}_{\ast}}
  \right)^{8/7}
  \left( \frac{m_{\rm H}^{2}}{\rho_{\ast}^{2} \Lambda_{\rm max}}
  \right)^{1/7}
  {f_{\ast}}^{4(1-\theta)/7}  \,\, .
  \label{eq:QTR}
\end{equation}
A potential roadblock to solving Equations (\ref{eq:rhoTR}--\ref{eq:QTR})
is that we do not initially know the value of $\tilde{\alpha}_{\rm TR}$.
This quantity depends on the reflection coefficient ${\cal R}$, which
depends on the Alfv\'{e}n speed $V_{\rm A}$ at the TR (see
Equation (\ref{eq:RefLow})), which in turn depends on the unknown
value of $\rho_{\rm TR}$.
In practice, we solve these equations iteratively.
We start with an initial estimate of ${\cal R} = 0.5$, we compute
$\tilde{\alpha}_{\rm TR}$, $\rho_{\rm TR}$, and $V_{\rm A}$ at the TR,
and then we recompute ${\cal R}$ for the next iteration.
In all cases the process converges to a self-consistent set of values
(with a relative accuracy of $\sim 10^{-7}$) in no more than 20
iterations.

The mass loss rate of the stellar wind is determined by the heating rate
$Q_{\rm TR}$ as well as other sources and sinks of energy at the TR.
The general idea that the solar wind's mass flux is set by the
energy balance at the TR was first discussed by \citet{Hm82}.
\citet{HL95} worked out the basic scaling argument that is used below
\citep[see also][]{Le82,Wi88,SM03}.
In the low corona and wind, the time-steady equation of internal energy
conservation is
\begin{equation}
  \frac{1}{A} \frac{\partial}{\partial r} \left\{ A \left[
  F_{\rm H} - F_{\rm cond} + \rho u \left( \frac{u^2}{2} -
  \frac{GM_{\ast}}{r} \right) \right] \right\} \, = \, 0 \,\, , 
  \label{eq:internal}
\end{equation}
where $F_{\rm H}$ is the energy flux associated with the heating,
$F_{\rm cond}$ is the energy flux transported by heat conduction
along the field, and $u$ is the outflow speed.
The term in braces is constant as a function of radius, so it is
straightforward to equate its value at the TR to its asymptotic
value at $r \rightarrow \infty$.
The kinetic energy term proportional to $u^2$ is assumed to be
negligibly small at the TR, but we assume it dominates the energy
balance at large distances.
Thus,
\begin{equation}
  A_{\rm TR} \left( F_{\rm H, TR} - F_{\rm cond} \right) -
  (\rho u A)_{\rm TR} \frac{GM_{\ast}}{R_{\ast}} \, = \,
  (\rho u A)_{\infty} \frac{u_{\infty}^2}{2}  \,\, ,
  \label{eq:Mbalance}
\end{equation}
where $F_{\rm H, TR}$ is the heat flux $F_{\rm H}$ at the TR, and
we realize that the product $\rho u A$ is also constant via
mass flux conservation.
We also make the key assumption that $u_{\infty} = V_{\rm esc} =
(2GM_{\ast} / R_{\ast})^{1/2}$, and thus we can write
\begin{equation}
  \dot{M} \, \equiv \, \rho u A \, = \,
  \frac{4\pi R_{\ast}^{2} f_{\rm TR}}{V_{\rm esc}^2}
  \left( F_{\rm H, TR} - F_{\rm cond} \right) \,\, .
  \label{eq:Mdothot}
\end{equation}

To evaluate Equation (\ref{eq:Mdothot}) we need to estimate the
value of $F_{\rm H, TR}$.
Formally, $Q = | \nabla \cdot {\bf F}_{\rm H} |$, so
to determine the magnitude $F_{\rm H, TR}$ one would have to
integrate $Q(r)$ along the flux tube.
Taking account of the expanding flux tube area $A \propto B^{-1}$,
and also assuming that $r_{\rm TR} \approx R_{\ast}$,
\begin{equation}
  F_{\rm H, TR} \, = \, \frac{1}{A_{\rm TR}} \int_{R_{\ast}}^{\infty}
  dr \, Q(r) \, A(r) \,\, .  
  \label{eq:FTRdef}
\end{equation}
Specifically, if $Q \propto r^{-\beta}$ and $A \propto r^{\gamma}$, then
\begin{equation}
  F_{\rm H, TR} \, = \, \frac{Q_{\rm TR} R_{\ast}}{|\beta - \gamma - 1|}
  \, \equiv \, Q_{\rm TR} R_{\ast} h \,\, .
  \label{eq:FTRemp}
\end{equation}
Rather than specifying $\beta$ and $\gamma$, we estimate the
dimensionless scaling factor $h$ by extracting both
$Q_{\rm TR}$ and $F_{\rm H, TR}$ from the self-consistent
solar wind models of \citet{CvB07}.
For a range of fast and slow solar wind solutions, we found that
$Q_{\rm TR}$ is typically between $1.5 \times 10^{-5}$ and
$4 \times 10^{-5}$ erg cm$^{-3}$ s$^{-1}$, and
$F_{\rm H, TR}$ is typically between $8 \times 10^{5}$ and
$3 \times 10^{6}$ erg cm$^{-2}$ s$^{-1}$.
This results in $h$ usually being between 0.5 and 1.5.

It is important to also verify that $F_{\rm H, TR}$ is less than
the energy flux carried ``passively'' by the Alfv\'{e}n waves as they
propagate up from the photosphere.
The latter quantity, which we call $F_{\rm A, TR}$, represents the
upper limit of available energy in the waves (at the TR) that can be
extracted by the turbulent heating.
Assuming that wave flux is conserved (i.e., that $M_{\rm A} \ll 1$
at the TR), then
$F_{\rm A, TR} = f_{\ast} F_{\rm A \ast} / f_{\rm TR}$.
For the cool-star models discussed in Section \ref{sec:results},
we found that the ratio $F_{\rm H, TR} / F_{\rm A, TR}$ is
usually around 0.1 to 0.5.
Only in two cases did it exceed 1 (albeit with values no larger than
1.5), and in those cases we capped $F_{\rm H, TR}$ to be equal
to $F_{\rm A, TR}$ to maintain energy conservation.

To evaluate Equation (\ref{eq:Mdothot}), we also need to estimate the
magnitude of the downward conductive flux $F_{\rm cond}$ from the
hot corona.
For the solar TR and low corona, \citet{Wi88} found there to be an
approximate balance between conduction and radiation losses.
\citet{Wi88} determined that
$F_{\rm cond} \approx c_{\rm rad} P_{\rm TR}$, where $P_{\rm TR}$ is
the gas pressure at the TR, and the constant of proportionality is
\begin{equation}
  c_{\rm rad} \, = \, \sqrt{\frac{\kappa_e}{2 k_{\rm B}^2} 
  \int_{T_0}^{T_{\rm TR}} \Lambda(T) \, T^{1/2} \, dT}
\end{equation}
where $\kappa_{e}$ is the electron thermal conductivity,
$T_{0} \approx 10^{4}$ K is a representative chromospheric temperature,
$T_{\rm TR} = 2 \times 10^{5}$ K, and $c_{\rm rad}$ has units of speed.
We evaluated the above integral to be able to scale out the
metallicity-dependent factor given in Equation (\ref{eq:Lammax})
above, and found that
\begin{equation}
  c_{\rm rad} \, \approx \, 14 \,
  \sqrt{\frac{\Lambda_{\rm max}(Z)}{\Lambda_{\rm max}(Z_{\odot})}}
  \,\,\,\, \mbox{km} \,\, \mbox{s}^{-1} \,\, .
  \label{eq:cradZ}
\end{equation}
We used this expression to estimate $F_{\rm cond}= c_{\rm rad} P_{\rm TR}$.
For the specific case of the Sun, conduction is relatively unimportant
in open flux tubes, since $F_{\rm cond} \lesssim 0.05 F_{\rm H, TR}$.
For the other stars modeled in this paper, the ratio
$F_{\rm cond} / F_{\rm H, TR}$ spanned several orders of
magnitude from $10^{-4}$ to $10^{-1}$.
In the eventuality that the estimated value of $F_{\rm cond}$ may
exceed $F_{\rm H, TR}$, one would need an improved description of the
coronal temperature $T(r)$ to compute a more accurate value of the
conduction flux.
In our numerical code, however, we do not allow $F_{\rm cond}$ to exceed
a value of $\xi F_{\rm H, TR}$, where $\xi$ is an arbitrary constant
that we fixed to a value of 0.9.
This condition was not met for any of the stars in the database of
Section \ref{sec:results}.

Once these energy fluxes are computed, we then compute $\dot{M}$
using Equations (\ref{eq:QTR}), (\ref{eq:Mdothot}), and
(\ref{eq:FTRemp}), as well as the other definitions given above.
Interestingly, this can be done without needing to know the
temperature profile $T(r)$.
From a certain perspective, the corona's thermal response to the
heating rate $Q$ may be considered to be just an intermediate step
toward the ``final'' outcome of a kinetic-energy-dominated outflow
far from the star.\footnote{%
Our approach, which ignores the details of this intermediate step,
is an approximation that also sidesteps some other important issues.
For example, a time-steady wind solution should pass through one or
more critical points, and it and should also satisfy physical boundary
conditions at $r = R_{\ast}$ and $r \rightarrow \infty$.
In Section~\ref{sec:conc} we summarize the necessary steps to
producing more self-consistent versions of this model.}
However, it should be possible to estimate the maximum coronal
temperature $T_{\rm max}$ by inverting scaling laws given by, e.g.,
\citet{Hm96} and \citet{SM03}.

The mass loss rate given by Equation (\ref{eq:Mdothot}) depends on
our assumption that $u_{\infty} = V_{\rm esc}$.
Equation (\ref{eq:Mbalance}) shows that larger assumed values of
$u_{\infty}$ would give rise to lower mass loss rates, and smaller
values of $u_{\infty}$ would give larger mass loss rates.
For the solar wind there is roughly a factor of three variation in
$u_{\infty}$, from about $0.4 V_{\rm esc}$ to $1.3 V_{\rm esc}$.
For other stars, it is rare to see observations where $u_{\infty}$
exceeds $V_{\rm esc}$, and \citet{Ju92} found generally that
$u_{\infty} < V_{\rm esc}$ for luminous evolved stars.
Even in the extreme case of $u_{\infty} = 0$, however,
Equation (\ref{eq:Mbalance}) would give only two times the mass loss
assumed by Equation (\ref{eq:Mdothot}).
When compared to the larger typical observational uncertainties in
$\dot{M}$, factors of two are not a major concern.

The Sun's mass loss rate of $2 \times 10^{-14}$ to
$3 \times 10^{-14}$ $M_{\odot}$ yr$^{-1}$ is modeled reasonably well
with the model described here.
The photospheric energy flux of Alfv\'{e}n waves is
$F_{\rm A \ast} \approx 1.5 \times 10^{8}$ erg cm$^{-2}$ s$^{-1}$, and
the photospheric wave amplitude is
$v_{\perp \ast} \approx 0.28$ km s$^{-1}$ \citep{CvB07}.
Although magnetogram observations sometimes give filling factors
$f_{\ast}$ as large as 1\% at solar maximum, values of 0.1\% tend to
better represent the coronal holes that are connected to the largest
volume of open flux tubes \citep[see Figure 3 of][]{CvB05}.
Assuming $f_{\ast} = 0.001$ and values for the other constants of
$\alpha_{0} = 0.5$ and $\theta = 1/3$,
Equations (\ref{eq:rhoTR}--\ref{eq:QTR}) give
$\rho_{\rm TR} \approx 5 \times 10^{-16}$ g cm$^{-3}$ and
$Q_{\rm TR} \approx 4 \times 10^{-5}$ erg cm$^{-3}$ s$^{-1}$, which
are in agreement with the models of \citet{CvB07} and others.
Thus, with $h = 0.5$, Equation (\ref{eq:Mdothot}) gives
$\dot{M} \approx 3.5 \times 10^{-14}$ $M_{\odot}$ yr$^{-1}$.

Although the above calculation of $\dot{M}$ is relatively
straightforward, it has not been boiled down to a simple scaling
law such as that of \citet{Rm75,Rm77}, \citet{Mu78}, or \citet{SC05}.
However, if we make the further assumptions that $\tilde{\alpha}$ and
$h$ are fixed constants, and that $F_{\rm H, TR} \gg F_{\rm cond}$,
we can isolate several interesting scalings:
\begin{enumerate}
\item
The ultimate driving of the wind comes from the basal flux of
Alfv\'{e}n wave energy $F_{\rm A \ast}$.
\citet{SC05} assumed that $\dot{M}$ scales linearly with
$F_{\rm A \ast}$, but in our case we can combine the above equations
with the definition of $Q_{\ast}$ to find
$\dot{M} \propto F_{\rm A \ast}^{12/7}$, which is noticeably
steeper than a pure linear dependence.
This {\em positive feedback} is qualitatively similar to what occurs
in radiatively driven winds of more massive stars, for which the mass
loss rate is proportional to the radiative flux (or luminosity) to a
power larger than one
\citep[i.e., $\dot{M} \propto L_{\ast}^{1.7}$;][]{CAK,Ow04}.
\item
Extracting the dependence on magnetic filling factor, we found that
$\dot{M} \propto f_{\ast}^{(4 + 3\theta)/7}$.
Using the range of $\theta$ from solar models (0.3--0.5), this
gives a relatively narrow range of exponents,
$\dot{M} \propto f_{\ast}^{0.7}$ to $f_{\ast}^{0.8}$.
\citet{Sa96a} estimated that $f_{\ast} \propto P_{\rm rot}^{-1.8}$
for rotation periods $P_{\rm rot} > 3$ days (see also
Section \ref{sec:mag}).
Thus, for stars in the unsaturated part of the age-activity-rotation
relationship, it may be possible to estimate
$\dot{M} \propto P_{\rm rot}^{-1.3}$.
\end{enumerate}
These simple scaling relations are given only for illustrative purposes
(see also Equation (\ref{eq:toosimp}) below).
The predictions of our ``hot'' coronal mass loss model should be
considered to be the solutions of the full set of
Equations (\ref{eq:lamstar}--\ref{eq:cradZ}).

\subsection{Cold Wave-Driven Mass Loss}
\label{sec:mdot:cold}

\begin{figure}
\epsscale{1.09}
\plotone{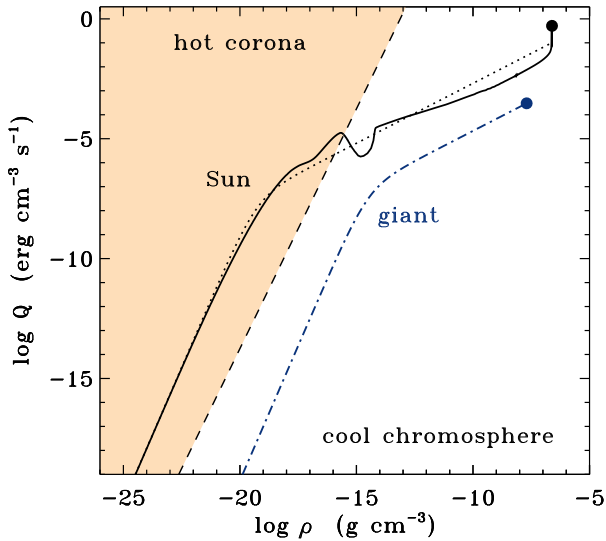}
\caption{Density dependence of the heating rate
$Q$ defined in Equation (\ref{eq:Qbasic}) compared to the
density dependence of the maximum radiative cooling rate from
Equation (\ref{eq:maxcool}) (dashed curve).
For the Sun, a numerical model (black solid curve) compares favorably
with a simple analytic ``bridging'' between the near-star
($Q \propto \rho^{1/2}$) and distant ($Q \propto \rho^{9/4}$)
scalings discussed in the text (black dotted curve).
For an example evolved giant star (blue dot-dashed curve), the
transition to the steeper density dependence occurs to the right of
the cooling boundary.
\label{fig04}}
\end{figure}

In a high-density stellar atmosphere, it is possible that the
turbulent heating described by Equation (\ref{eq:Qbasic}) could
be balanced by radiative cooling even very far from the star.
In that case, hot coronal temperatures may never occur
\citep[see, e.g.,][]{Su07,Cr08}.
The density dependence of the heating rate $Q$ determines whether
radiative cooling remains important at large distances, and
Figure \ref{fig04} shows two examples of how $Q$ may vary as a
function of $\rho$.
Near the stellar surface, where $v_{\perp} \propto \rho^{-1/4}$
and $B \propto \rho^{1/2}$, we can combine various assumptions to
estimate that $Q \propto \rho^{1/2}$.
However, further from the star, where $v_{\perp} \propto \rho^{+1/4}$
and $B \propto \rho$, the density dependence becomes steeper, with
$Q \propto \rho^{9/4}$.

Figure \ref{fig04} compares the modeled heating rates with the
``maximum cooling boundary'' implied by Equation (\ref{eq:maxcool}).
The solar model crosses the boundary, and thus undergoes a
transition to a hot corona.
The solid curve was taken from a numerical model of fast solar wind
from a polar coronal hole \citep{CvB07}.
On the other hand, a model for a representative late-type giant sits
to the right of the cooling boundary, which implies that radiative
losses can maintain the circumstellar temperature at chromospheric
values of $\sim 10^{4}$ K even at large distances.
The two models differ because the density at which
$M_{\rm A} \approx 1$ (where the $Q$ curves undergo a change in
slope) for the giant is several orders of magnitude larger than the
corresponding density for the solar case.
This density is an output of a given mass loss model and cannot be
specified a~priori.

In this section we develop a model for cool-star mass loss under
the assumption of strong radiative cooling.
In this case the \citet{P58} gas pressure driving mechanism cannot
drive a significant outflow.
However, when the flux of Alfv\'{e}n waves is large, they can impart
a strong bulk acceleration to the plasma due to {\em wave pressure,}
which is a nondissipative net ponderomotive force exerted by virtue
of wave propagation through an inhomogeneous medium \citep{BG68,J77}.
The subsequent calculation of $\dot{M}$ for a ``cold wave-driven''
stellar wind largely follows the development of \citet{He83}
\citep[see also][]{Cr09}.

Three key assumptions are: (1) that the Alfv\'{e}n wave amplitudes
in the wind are larger than the local sound speeds,
(2) that there is negligible wave damping between the stellar
surface and the wave-modified critical point of the flow,
and (3) that the critical point occurs far enough from the stellar
surface that $f \approx 1$ there (i.e., the flux tube expansion
becomes radial).
A fourth assumption from \citet{He83}---which was initially
not applied here but later found to be valid---is that the stellar
wind is sub-Alfv\'{e}nic at the critical point (i.e., that
$M_{\rm A} \ll 1$ at the critical point).
\citet{Cr09} constructed a set of numerical models for the
cold polar outflows of T~Tauri stars that did not make the
fourth assumption, and found that for a wide range of parameters
the assumption was justified.
Thus, here we use the value of the critical radius given by
Equation (35) of \citet{He83}, in which all four of the above
assumptions were applied, and
\begin{equation}
  \frac{r_{\rm crit}}{R_{\ast}} \, \approx \,
  \frac{7/4}{1 + (v_{\perp \ast} / V_{\rm esc})^2} \,\, .
  \label{eq:rcrit}
\end{equation}

Once the critical point radius is known, it becomes possible to use
the known properties of the Alfv\'{e}n waves to determine the wind
velocity and density at the critical point.
\citet{He83} found analytic solutions for these quantities in
the limiting case of $M_{\rm A} \ll 1$ at the critical point.
Here we describe a slightly more self-consistent way of computing
$M_{\rm A}$ and the mass loss rate, but we also continue to use
Equation (\ref{eq:rcrit}) that was derived in the limit of
$M_{\rm A} \ll 1$.
There are three unknown quantities and three equations to
constrain them.
The three unknowns are the critical point values of the wind
speed $u$, density $\rho$, and wave amplitude $v_{\perp}$.
The first equation is the constraint that the right-hand side of
the time-steady momentum equation must sum to zero at the
critical point of the flow \citep[e.g.,][]{P58}.
For the conditions described above, this gives
\begin{equation}
  \frac{2 u_{\rm crit}^{2}}{r_{\rm crit}} -
  \frac{GM_{\ast}}{r_{\rm crit}^2} \, = \, 0 \,\, ,
\end{equation}
and it is solved straightforwardly for $u_{\rm crit}$.
The second and third equations are, respectively, the definition
of the critical point velocity in the ``cold'' limit of zero
gas pressure,
\begin{equation}
  u_{\rm crit}^{2} \, = \, \frac{v_{\perp}^2}{4}
  \left( \frac{1 + 3 M_{\rm A}}{1 + M_{\rm A}} \right)
\end{equation}
and the conservation of wave action as given by
Equation (\ref{eq:action}).
The fact that $V_{\rm A}$ appears in these equations and depends
on the (still unknown) density makes it difficult to find an
explicit analytic solution for $\rho_{\rm crit}$.
We again use iteration from an initial guess to reach a self-consistent
solution for $u$, $\rho$, and $v_{\perp}$ at the critical point.
The stellar wind's mass loss rate is thus determined from
$\dot{M} = 4\pi r_{\rm crit}^{2} u_{\rm crit} \rho_{\rm crit}$.

Because the mass loss rate is set at the critical point, we do not
need to specify the terminal speed $u_{\infty}$.
For most implementations of the above model, the denominator in
Equation (\ref{eq:rcrit}) is close to unity and thus we have 
$u_{\rm crit}^{2} \approx V_{\rm esc}^{2} / 7$, or that
$u_{\rm crit}$ is about 38\% of the presumed value of $u_{\infty}$.
Of course, there have been stellar wind models with nonmonotonic
radial variations of $u(r)$, with $u_{\rm crit} > u_{\infty}$
\citep[e.g.,][]{FG06}.
It is also possible for ``too much'' mass to be driven past the
critical point, such that parcels of gas may be decelerated to
stagnation at some height above $r_{\rm crit}$ and thus would want
to fall back down towards the star.
In reality, this parcel would collide with other parcels that
are still accelerating, and a stochastic collection of shocked
clumps is likely to result.
Interactions between these parcels may result in an extra degree
of collisional heating that could act as an extended source of
gas pressure to help maintain a mean net outward flow.
Situations similar to this have been suggested to occur in the
outflows of pulsating cool stars \citep{Bw88,Sr04}, T Tauri
stars \citep{Cr08}, and luminous blue variables \citep{vM09}.

\subsection{Combining Hot and Cold Models}
\label{sec:mdot:both}

A proper treatment of a stellar wind powered by MHD
turbulence---and accelerated by a combination of gas pressure and
wave pressure effects---requires a self-consistent
numerical solution to the conservation equations
\citep[e.g.,][]{CvB07,Su07,Co09,Ai10}.
However, in this paper, we explore simpler ways of estimating the
combined effects of both processes.

Sections \ref{sec:mdot:hot} and \ref{sec:mdot:cold} gave us independent
estimates for the mass loss rate assuming only gas pressure or wave
pressure were active in the flux tube of interest.
We refer to these two mass loss rates as $\dot{M}_{\rm hot}$ and
$\dot{M}_{\rm cold}$, respectively.
It seems clear that when one of these values is much larger than
the other, then one process is dominant and the actual mass loss
rate should be close to that larger value.
For the manifestly ``hot'' example of the Sun, we found that
$\dot{M}_{\rm hot} / \dot{M}_{\rm cold} \approx 20$, which correctly
implies that gas pressure driving is dominant.
For most examples of late-type giants with
$L_{\ast} > 10 \, L_{\odot}$, the ratio
$\dot{M}_{\rm hot} / \dot{M}_{\rm cold}$ was found to decrease to
values between about 0.1 and 3.
This could mean that gas and wave pressure gradients are of the same
order of magnitude for these stars.

One of the most straightforward things that can be done is to assume
the combined effect of gas and wave pressure produces a
mass loss rate equal to the {\em sum} of the two individual components,
$\dot{M}_{\rm hot} + \dot{M}_{\rm cold}$.
This preserves the idea that one dominant mechanism should determine
$\dot{M}$ when the other would predict a negligibly small effect.
It also makes sense based on Equation (\ref{eq:internal}), which
shows how the energy fluxes sum together linearly in the internal
energy equation.
If there were multiple sources of input energy flux,
Equation (\ref{eq:Mdothot}) would show that the resulting mass
loss rate should be proportional to their sum.

However, there is one complication that hinders us from simply
adding together $\dot{M}_{\rm hot}$ and $\dot{M}_{\rm cold}$.
The calculation of $\dot{M}_{\rm hot}$ from
Section \ref{sec:mdot:hot} contains the assumption that the
TR turbulent heating always obeys the near-star density scaling
$Q \propto \rho^{1/2}$.
It therefore predicts that all stars eventually undergo a transition
to a hot corona.
For some stars (like the late-type giant in Figure \ref{fig04}),
however, we know that there should be no corona and it is
erroneous to assume that $\dot{M}_{\rm hot}$ has any real meaning.
Thus, for each model we compute the wind speed at the TR from
mass flux conservation,
\begin{equation}
  u_{\rm TR} \, = \, \frac{\dot{M}_{\rm hot}}
  {4\pi R_{\ast}^{2} f_{\rm TR} \rho_{\rm TR}}  \,\, ,
\end{equation}
and we demand that for $\dot{M}_{\rm hot}$ to have a consistent
interpretation, the TR Mach number $M_{\rm A, TR} = u_{\rm TR} /
V_{\rm A, TR}$ should be much smaller than one.
As expected, this condition was found to be violated for late-type
giants having $L_{\ast} \gtrsim 100 \, L_{\odot}$.
Thus, in these cases we should replace $\dot{M}_{\rm hot}$ with either
a drastically reduced value or zero---the latter in cases where the
$Q(\rho)$ curve always falls to the right of the maximum cooling
boundary in Figure \ref{fig04}.
After some experimentation, we found that reducing the initially
computed value of $Q_{\rm TR}$ by a factor of
$\exp ( - 4 M_{\rm A, TR}^{2})$ does a reasonably good job of
reproducing the result of using a more consistent $Q(\rho)$
function.
Thus, we propose that the summing of the ``hot'' and ``cold''
mass loss rates be done with the following approximate expression,
\begin{equation}
  \dot{M} \, \approx \, \dot{M}_{\rm cold} +
  \dot{M}_{\rm hot} \exp \left( - 4 M_{\rm A, TR}^{2} \right)
  \label{eq:combine}
\end{equation}
where $\dot{M}_{\rm hot}$ and $M_{\rm A, TR}$ are computed using
the assumptions of Section \ref{sec:mdot:hot} and
$\dot{M}_{\rm cold}$ is computed using the model given in
Section \ref{sec:mdot:cold}.

\section{Magnetic Activity and Rotation}
\label{sec:mag}

An important ingredient in the above models---which remains unspecified
for most stars---is the photospheric filling factor $f_{\ast}$.
It is now well-known that both $f_{\ast}$ and the magnetic flux
density $B_{\ast} f_{\ast}$ exhibit significant correlations with
stellar rotation speed \citep{SL86,MB89,MJ93,Sa01}.
For many stars the rotation rate also scales with age, chromospheric
activity, and coronal X-ray emission \citep{Sk72,Ny84,Pz03,MH08}.
A prevalent explanation for these correlations is that an MHD dynamo
amplifies the magnetic flux in proportion to the large-scale energy
input from differential rotation
\citep[e.g.,][]{P79,Mo01,Bu03,MS09,Cu09,Is11}.

In this section we construct an empirical scaling relation that will
allow a reasonably accurate determination of $f_{\ast}$ as a function
of $P_{\rm rot}$ and the other stellar parameters.
Other estimates of this relationship have been made in the past
\citep{MJ93,St94,Sa96a,Sa01,Cu98,Fw02}.
However, since our goal is to apply this relation to stellar wind
acceleration (in open flux tubes that cover a subset of the inferred
$f_{\ast}$ area) and to evolved giants (which are greatly undersampled
in observational studies of $f_{\ast}$), we
aim to reanalyze the existing data rather than rely on other
published scalings.

Table 1 lists the properties of 29 stars that have
reliable measurements of their fundamental parameters, rotation
rates, and either independent or combined values of $B_{\ast}$
and $f_{\ast}$.
The sources for these values are given as numbered references in the
final column.
In many cases the available sources gave only a subset of the basic
stellar parameters.
When necessary, we used Equation (\ref{eq:basic}) and information
from the NASA/IPAC/NExScI Star and Exoplanet Database (NStED)%
\footnote{\url{http://nsted.ipac.caltech.edu/}} to fill in missing
values \citep[see][]{Br10}.
Table 1 also gives approximate ``quality factors'' $q$
that describe the relative accuracy of the measurements, and
in the Appendix we describe these factors in more detail.
\begin{figure*}
\epsscale{1.13}
\plotone{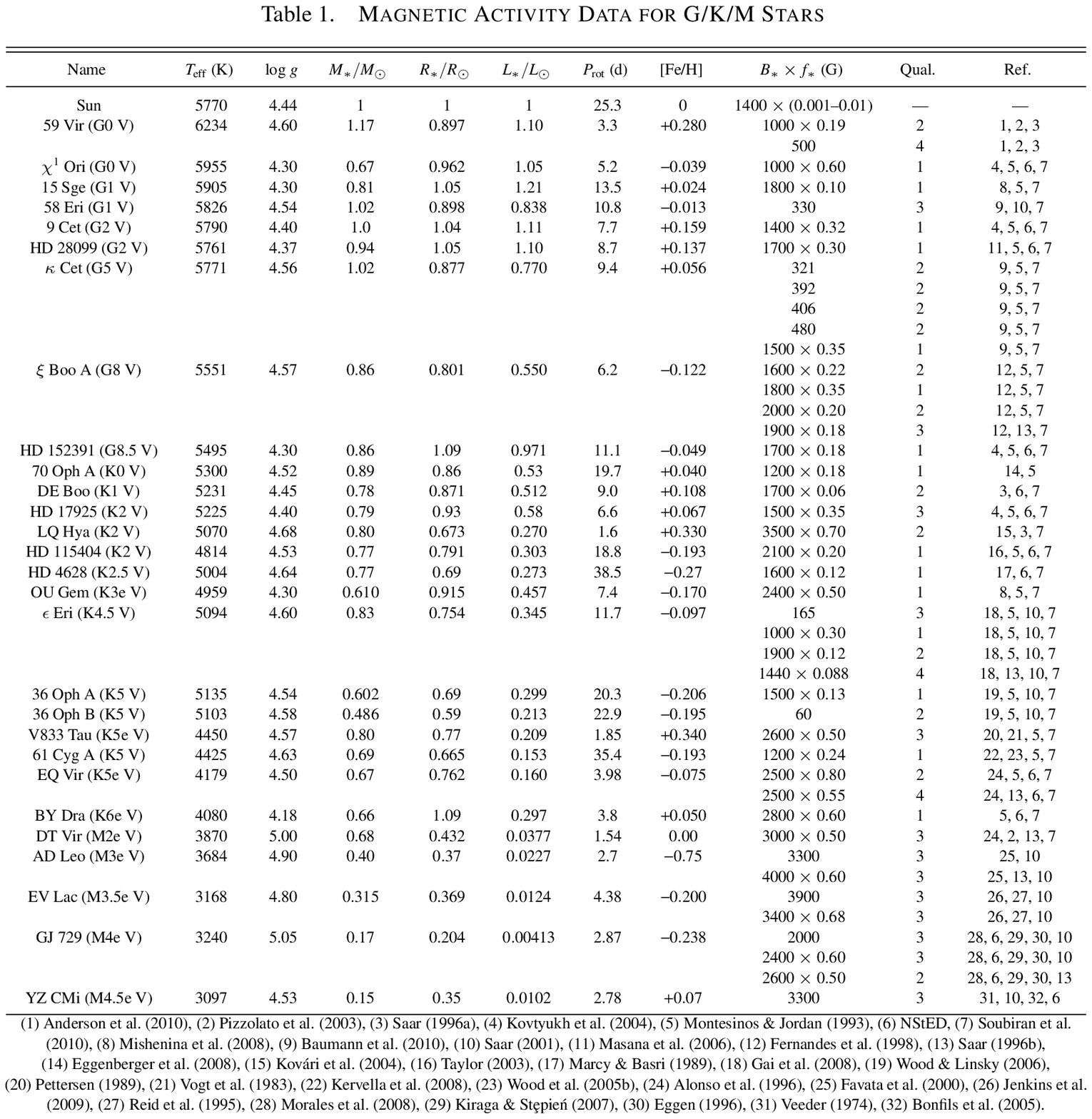}
\end{figure*}

\begin{figure}
\epsscale{1.09}
\plotone{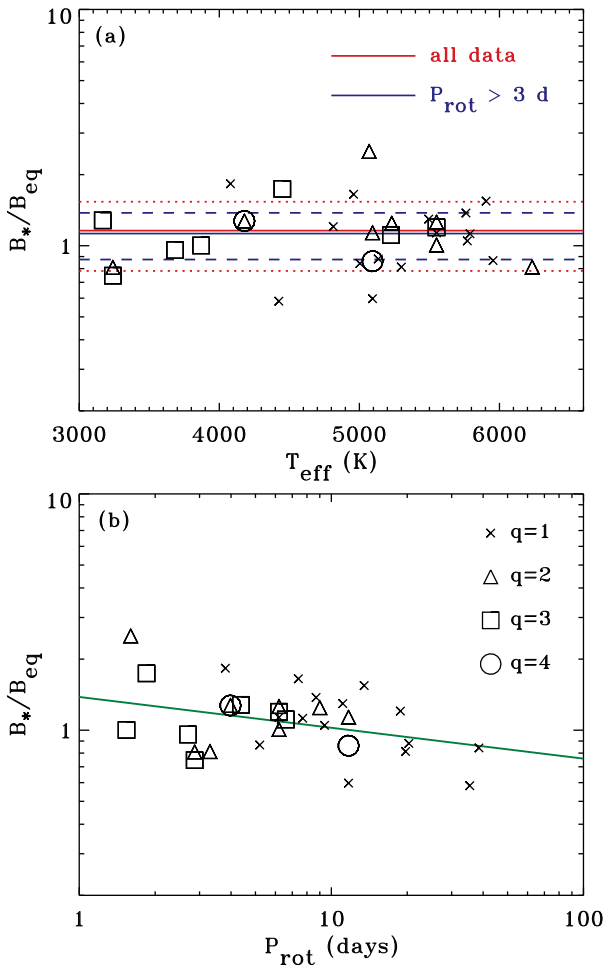}
\caption{Observational data for $B_{\ast}/B_{\rm eq}$ as a
function of (a) effective temperature and (b) rotation rate.
Solid lines in (a) show mean values for all data (red) and for only
stars having $P_{\rm rot} > 3$ days (blue).
Dotted and dashed lines show regions within $\pm 1 \sigma$ of the means.
Quality factors are denoted by crosses ($q=1$), triangles ($q=2$),
squares ($q=3$) and circles ($q=4$).
\label{fig05}}
\end{figure}

It has been known for some time that, for dwarf stars, $B_{\ast}$
never appears to be very far from the equipartition field strength
$B_{\rm eq}$ \citep[e.g.,][]{SL86}.
Figure \ref{fig05} plots the ratio $B_{\ast}/B_{\rm eq}$ for the
measurements in Table 1 that have separate determinations
of $B_{\ast}$ and $f_{\ast}$.
The sizes of the symbols are proportional to the observational
quality factors, and all statistical fits and moments discussed
below were weighted linearly with $q$.
Figure \ref{fig05}(a) shows that there is no strong correlation of
$B_{\ast}/B_{\rm eq}$ with $T_{\rm eff}$.
\citet{Sa96a} found a slight increase in $B_{\ast}/B_{\rm eq}$
for the most rapid rotators ($P_{\rm rot} < 3$ days), and
Figure \ref{fig05}(b) shows that when more data are included this
trend survives but is not strong.
The power-law fit is consistent with a relationship
$B_{\ast} \propto P_{\rm rot}^{-0.13}$, but we do not consider it
significant enough to apply it below or to extrapolate it to longer
rotation periods.

We found that the $q$-weighted mean value of $B_{\ast}/B_{\rm eq}$
for the entire sample (1.16, with standard deviation $\pm 0.38$) is
only marginally higher than the mean value for the subset of
slower rotating, non-saturated stars with $P_{\rm rot} > 3$ days
(1.13, with standard deviation $\pm 0.25$).
Equation (\ref{eq:Bstar}) gives the latter mean value, which we use
in Section \ref{sec:results} for modeling the winds of the
(generally slowly rotating) stars with observed mass loss rates.
We also use Equation (\ref{eq:Bstar}) to estimate $f_{\ast}$
for the cases where only the product $B_{\ast} f_{\ast}$ has been
measured.

A primary indicator of stellar magnetic activity appears to be
the photospheric filling factor $f_{\ast}$.
There have been a number of different proposed ways to express
the general anticorrelation between activity and rotation period.
\citet{Ny84} found that indices of chromospheric activity
correlate better with the so-called {\em Rossby number}
$\mbox{Ro} \equiv P_{\rm rot}/ \tau_{c}$, where $\tau_c$ is a
measure of the convective turnover time, than with
$P_{\rm rot}$ alone.
For other data sets, however, the usefulness of the Rossby number
has been called into question \citep{Bs86,St94}.
\citet{Sa91} postulated that $B_{\ast} f_{\ast}$ (and presumably
also $f_{\ast}$ itself) is proportional to $\mbox{Ro}^{-1}$
\citep[see also][]{MJ93,Cu98,Sa01,Fw02}.

\begin{figure}
\epsscale{1.09}
\plotone{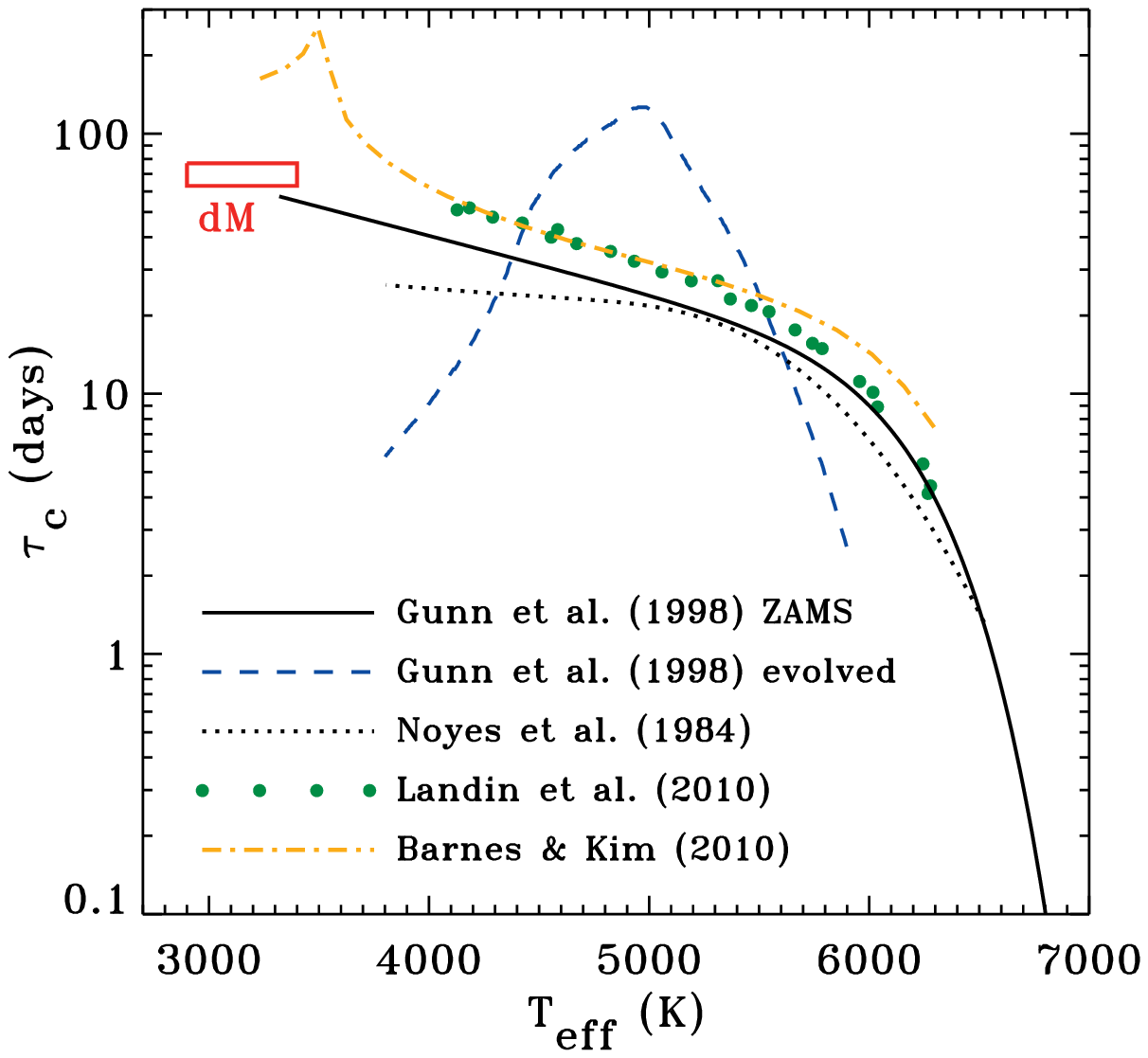}
\caption{Estimates of the convective turnover time $\tau_c$
as a function of $T_{\rm eff}$.
ZAMS models of \citet{Gu98} were fit by Equation (\ref{eq:gunn})
(black solid curve).
The \citet{Gu98} evolutionary track for a $2.2 M_{\odot}$ star
(blue dashed curve) outlines an upper limit for $\tau_c$ at
intermediate temperatures.
Local values of $\tau_c$ from \citet{Ln10} (green symbols)
and \citet{BK10} (orange dot-dashed curve),
the parameterization given by \citet{Ny84} (black dotted curve),
and the M dwarf estimate of $\tau_{c} \approx 70$~d used by
\citet{Rn09} (red box labeled by ``dM'') are also shown.
\label{fig06}}
\end{figure}

To compute the Rossby number for a given star, we need to know the
convective turnover time $\tau_c$.
Figure \ref{fig06} compares several past calculations of
$\tau_{c}$ with one another.
For most stars we will utilize a parameterized fit to the set of
ZAMS stellar models given by \citet{Gu98},
\begin{equation}
  \tau_{c} = 314.24
  \exp \left[ - \left( \frac{T_{\rm eff}}{1952.5 \, \mbox{K}} \right)
  - \left(\frac{T_{\rm eff}}{6250 \, \mbox{K}} \right)^{18}
  \right] + 0.002 \, ,
  \label{eq:gunn}
\end{equation}
where $\tau_c$ is expressed in units of days and the fit is valid
for the approximate range $3300 \lesssim T_{\rm eff} \lesssim 7000$ K.
Such a fit ignores how $\tau_{c}$ may depend on other stellar parameters
besides effective temperature, but more recent sets of models
\citep{Ln10,BK10,KO11} also found reasonably monotonic behavior as a
function of $T_{\rm eff}$ for a broad range of stellar ages and masses.

There are indications that the simple relationship between $\tau_c$
and $T_{\rm eff}$ seen for main-sequence stars is not universal.
For example,
\begin{enumerate}
\item
Low-mass M dwarfs (with $M_{\ast} \lesssim 0.35 \, M_{\odot}$)
are likely to be fully convective, and thus their dynamos are likely
to be driven by fundamentally different processes than exist in
more massive stars \citep{MM01,RB07,Ir11}.
There is also some disagreement about the relevant $\tau_c$ values
for these stars.
Figure \ref{fig06} shows that the models of \citet{BK10} exhibit a
slight discontinuity at the fully convective boundary.
The \citet{Rn09} semi-empirical estimate of $\tau_{c} \approx 70$
days for M dwarfs is about a factor of 2--3 lower than that of
\citet{BK10}.
However, because the \citet{Rn09} value is in reasonable
agreement with an extrapolation of Equation (\ref{eq:gunn})
to lower effective temperatures, we will just use this expression
and not make any special adjustments to the Rossby numbers of
fully convective M dwarfs.
\item
Luminous evolved giants exhibit qualitatively different interior
properties than do main sequence stars of similar $T_{\rm eff}$.
Despite not having firm measurements of the magnetic activities of
evolved giants, we will want to estimate $f_{\ast}$ for such
stars in order to compute their mass loss rates.
\citet{Go05,Go07} found that the correlation between X-ray activity
and rotation in G and K giants is consistent with that of
main-sequence stars if the {\em larger} values of $\tau_{c}$ from
the evolved models of \citet{Gu98} were used instead of the ZAMS
values (see the blue dashed curve in Figure \ref{fig06}).
Similarly, \citet{Ha94} calculated luminosity-dependent scaling
factors that can be used to multiply the ZAMS value of $\tau_c$
to obtain a consistent relation between rotation and photometric
activity \citep[see also][]{Ch95}.
We found that the above results can be generally reproduced by
multiplying the ZAMS value of $\tau_c$ by a factor
$(g_{\odot}/g)^{0.23}$, which applies only for low-gravity
subgiants and giants (i.e., only for $g < g_{\odot}$).
In Section \ref{sec:results} we explore the extent to which this
kind of approximate correction factor helps to explain the activity
and mass loss of evolved stars.
\end{enumerate}
A slightly different way of estimating the magnetic flux of a
rotating star is to take advantage of a proposed
``magnetic Bode's law;'' i.e., the conjecture that the star's
magnetic moment scales linearly with its angular momentum
\citep{Ar95,Ba96}.
Using the stellar parameters defined above, this corresponds
approximately to
\begin{equation}
  B_{\ast} f_{\ast} R_{\ast}^{3} \, \propto \,
  \frac{M_{\ast} R_{\ast}^2}{P_{\rm rot}}  \,\,\, .
  \label{eq:bode}
\end{equation}
The above relationship does not take into account variations of
the moment of inertia (for different stars) away from an idealized
scaling of $I \sim M_{\ast} R_{\ast}^{2}$, and it assumes the
magnetic moment is dominated by a large-scale dipole component.
It is possible to test this idea with the data given in
Table 1 by evaluating the correlation between
$f_{\ast}$ and
$M_{\ast} (R_{\ast} P_{\rm rot} B_{\ast})^{-1}$.

\begin{figure} 
\epsscale{1.11}
\plotone{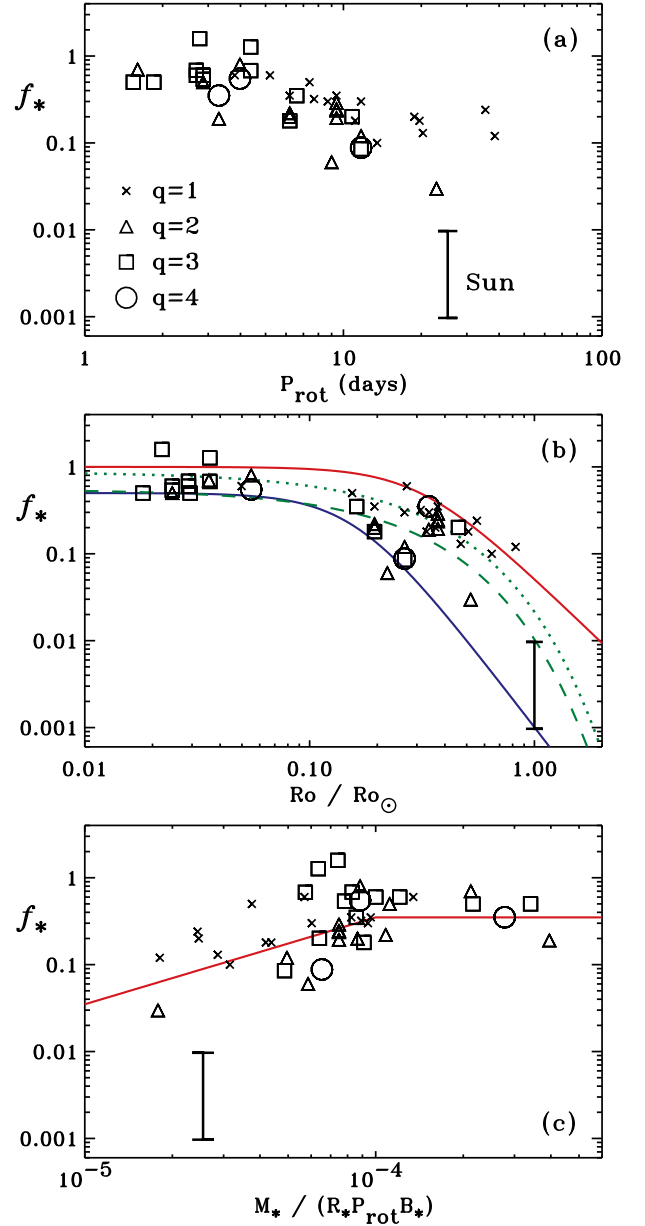}
\caption{Comparison of possible correlations between measured
$f_{\ast}$ filling factors with (a) rotation rate,
(b) Rossby number, and (c) a magnetic Bode's law parameter
(see Equation (\ref{eq:bode})).
Solid curves in (b) denote the lower (blue) and upper (red)
envelopes surrounding the data, and green curves show fitting
formulae from Equations 2.3 (dotted; empirical) and 7.3
(dashed; theoretical) of \citet{MJ93}.
Quality factors are denoted by the same symbols used in
Figure \ref{fig05}, and the Sun's range of $f_{\ast}$ is
shown with a vertical bar.
\label{fig07}}
\end{figure}

Figure \ref{fig07} shows how the empirical set of $f_{\ast}$
values correlates with rotation period, Rossby number, and the
proposed magnetic Bode's law.
Rather than use the Rossby number itself, we instead plot the data
in Figure \ref{fig07}(b) as a function of a normalized ratio
$\mbox{Ro} / \mbox{Ro}_{\odot}$, where according to
Equation (\ref{eq:gunn}), the Sun's Rossby number
$\mbox{Ro}_{\odot} = 1.96$.
Such a normalization allows us to neglect any scaling discrepancies
between ``local'' and ``global'' definitions of $\tau_c$
\citep[e.g.,][]{Pz01,Ln10}.
The Sun's large range of measured $f_{\ast}$ values ($10^{-3}$ to
$10^{-2}$) is indicated with a vertical bar, and it is likely that
all other stars exhibit such a range on both rotational and
dynamo-cycle time scales.

Figures \ref{fig07}(a) and \ref{fig07}(c) show that the
correlations with $P_{\rm rot}$ and the proposed magnetic Bode's
law are not especially strong.
However, if all of the lowest quality ($q=1$) measurements were
removed, the correlation with $P_{\rm rot}$ would be improved
significantly.
Figure \ref{fig07}(b) shows that the Rossby number seems to be a
slightly better ordering parameter, and it compares the individual
data points with several functional relationships.
The blue and red solid curves are subjective fits to the minimum
and maximum bounds on the envelope of data points, with
\begin{equation}
  f_{\rm min} \, = \, \frac{0.5}
  {[1 + (x / 0.16)^{2.6}]^{1.3}}  \,\, ,
  \label{eq:fmin}
\end{equation}
\begin{equation}
  f_{\rm max} \, = \, \frac{1}{1 + (x / 0.31)^{2.5}}
\end{equation}
where $x = \mbox{Ro} / \mbox{Ro}_{\odot}$.
We also show
empirical and theoretical fitting formulae from \citet{MJ93}.
Other comparisons could also be made with relationships given by
\citet{Cu98}, \citet{Fw02}, and others, but they all appear to
fall near the green and red curves.

Note that for slow rotation rates (i.e., large Rossby numbers) the
scaling laws shown in Figure \ref{fig07}(b) imply a significantly
{\em steeper} decline of $f_{\ast}$ than has been suggested in the past.
For example, \citet{Sa91} estimated $f_{\ast} \propto \mbox{Ro}^{-1}$,
and \citet{Sa96a} estimated $f_{\ast} \propto P_{\rm rot}^{-1.8}$.
On the other hand, our empirical upper and lower bounds suggest
$f_{\rm max} \propto \mbox{Ro}^{-2.5}$ and
$f_{\rm min} \propto \mbox{Ro}^{-3.4}$ respectively.
This is similar to the observed relationship between Rossby number
and X-ray activity.
\citet{MH08} found that the ratio of X-ray to bolometric luminosity
$L_{\rm X}/L_{\rm bol}$ drops by about a factor of 700 as the
Rossby number increases by a factor of ten from 0.25 to 2.5
\citep[see also][]{Wr11}.
This corresponds very roughly to a power-law decrease of
$\mbox{Ro}^{-2.85}$.
Its agreement with the behavior of $f_{\ast}$ shown above is
also consistent with existing empirical correlations between
X-rays and magnetic activity \citep{Pv03}.

In addition to the rotational scaling of $f_{\ast}$ with Ro, there
is also likely to be a ``basal'' lower limit on the outer atmospheric
activity of a star \citep[e.g.,][]{Sj87,Cu99,Bc05,TT11,PM11}.
Whether this lower limit is the result of acoustic waves, a
turbulent dynamo, or some other physical process, there is
probably a minimum value of $f_{\ast}$ that is independent
of rotation rate.
For example, \citet{Bc05} found that turbulent dynamos in main
sequence stars can generate a flux density 
$B_{\ast} f_{\ast} \approx 4$ G without much variation from
spectral types F0 to M0.
Using Equation (\ref{eq:Bstar}) for $B_{\ast}$, we can estimate a
basal filling factor for these stars of $f_{\ast} \approx 0.001$--0.002,
which is close to the Sun's minimum value.
However, since it is still uncertain whether or not the Sun has
exhibited truly basal flux conditions in recent years \citep{CL11},
we will set a slightly lower value of $f_{\rm basal} = 10^{-4}$
to be used in the mass loss models below.

Before moving on to apply the empirical values of $f_{\ast}$ to
our model of mass loss, we emphasize that the measurements do
not directly provide the filling factor of open magnetic flux tubes.
Ideally, Zeeman broadening measurements should be sensitive to the
{\em total} flux in strong magnetic elements on the stellar
surface, no matter whether the field lines are closed or open.
In many cases, however, the closed-loop active regions have
significantly stronger local field strengths than the open regions.
Therefore the closed-field regions are likely to dominate the
spectral line broadening that gives rise to the observational
determinations of $f_{\ast}$ (see the Appendix).
Without spatially resolved magnetic field measurements, we do not
yet have a definitive way to predict how a given star divides up
its flux tubes between open and closed.
\citet{MS87} claimed that as the rotation rate increases (from slow
values similar to the Sun's), the relative fraction of closed field
regions should first increase, then eventually it should decrease as
centrifugal forces strip the field lines open.
We can speculate that the {\em spread} in the measured $f_{\ast}$
data may tell us something about the closed and open fractions.
Because closed-loop active regions tend to have stronger fields
than open coronal holes, the lower and upper envelopes that
surround the data in Figure \ref{fig07}(b) could be good proxies for
the filling factors of open and closed regions, respectively.
More specifically, we hypothesize that
$f_{\rm min}$ is seen when no active regions are present on the
visible surface (i.e., $f_{\rm min} \approx f_{\rm open}$)
and that $f_{\rm max}$ is seen when active regions dominate the
observed magnetic flux.
This idea is tested, in a limited way, in
Section \ref{sec:results:compare}.

\section{Results}
\label{sec:results}

Here we present the results of solving the mass loss
equations derived in Section \ref{sec:mdot} using the empirical
estimates for the rotational dependence of the magnetic filling
factor derived in Section \ref{sec:mag}.
For hot coronal mass loss, we assumed values for the dimensionless
parameters $\alpha_{0} = 0.5$, $h = 0.5$, and $\theta = 1/3$.
As discussed above, these values were ``calibrated'' from our more
detailed knowledge of the Sun's coronal heating and wind acceleration.
Our use of these values for other stars is an extrapolation that
can be tested by comparison with observed mass loss rates.

\subsection{Database of Stellar Mass Loss Rates}
\label{sec:results:data}

\begin{figure*}
\epsscale{1.13}
\plotone{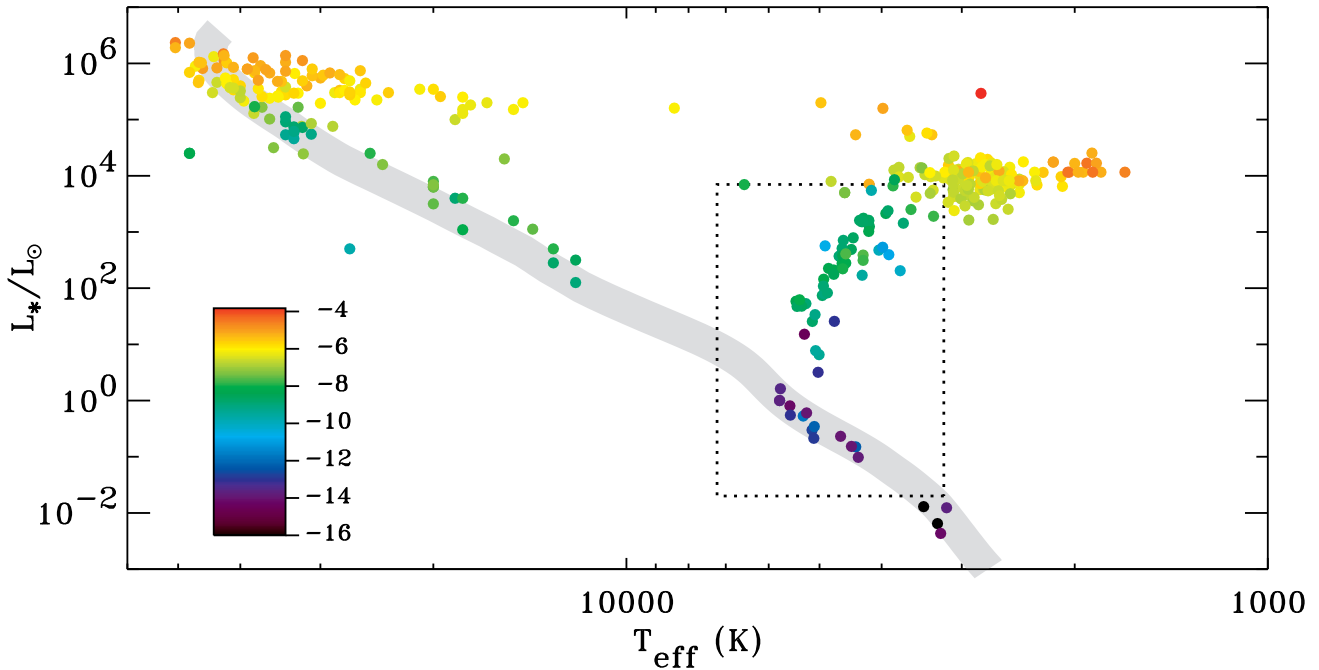}
\caption{Hertzsprung-Russell diagram showing observed mass loss rates.
The color scale at lower-left specifies log~$\dot{M}$, where $\dot{M}$
is measured in $M_{\odot}$ yr$^{-1}$.
An estimate for the ZAMS is shown in gray, and the approximate domain
of parameter space covered by the models of this paper is outlined
by a black dotted box.
\label{fig08}}
\end{figure*}

Figure \ref{fig08} is a broad overview of observed stellar mass loss.
It plots the locations of individual stars in a Hertzsprung-Russell
type diagram with their mass loss rates shown as symbol color
\citep[see also][]{dJ88}.
A box illustrates the approximate regime of parameter space covered
by the models developed in this paper; it extends from the
main sequence up through the regime of the so-called
``hybrid chromosphere'' stars \citep{Hy80}, and possibly
also into the parameter space of cool luminous supergiants.
In addition to the cool-star data discussed below, we also
include in Figure \ref{fig08} measured mass loss rates of
hot, massive stars \citep{Wa87,Lm99,Mk07,Se08},
FGK supergiants \citep{dJ88}, AGB stars \citep{Bg05,Gn10},
red giants in globular clusters \citep{Ms09},
and M dwarfs in pre-cataclysmic variable binaries \citep{Db06}.
Many of these stars are not included in the subsequent analysis
because we have no firm rotation periods or magnetic activity
indices for them.

Table 2 lists the properties of 47 stars for which
our knowledge appears to be complete enough to be able to
compare theoretical and observed values of $\dot{M}$.
For the Sun, the range of volume-integrated mass loss rates comes
from \citet{Wa98}.
The sources for all listed values are given as numbered references
that continue the sequence started in Table 1; the
citations corresponding to numbers 1--32 are given in
Table 1.
In cases where $T_{\rm eff}$, $\log \, g$, or [Fe/H] were
estimated from the PASTEL database \citep{PASTEL}, we averaged
together multiple measurements when more than one was given.
In the few cases where the same star appears in both
Table 1 and Table 2, for consistency's sake
we will recompute $f_{\ast}$ from the star's rotation period when
calculating theoretical mass loss rates
(see Section \ref{sec:results:compare}).
\begin{figure*}
\epsscale{1.13}
\plotone{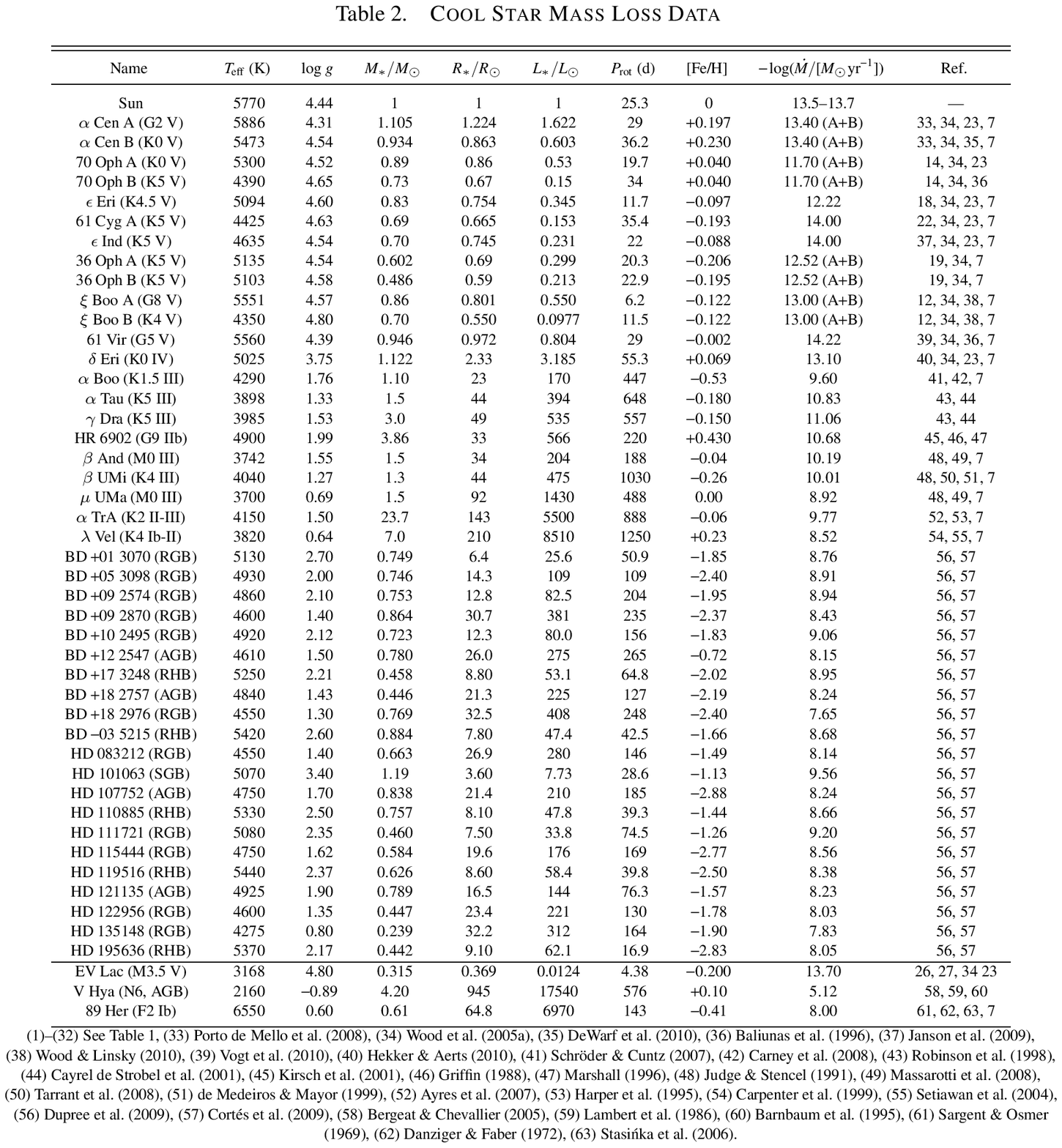}
\end{figure*}

For binary systems with astrospheric measurements of $\dot{M}$
\citep[see, e.g.,][]{Wo02}, the numbers given are assumed to be
the sum of both stars' mass loss rates.
We list that same value for both components and denote it
with ``(A$+$B).''
We did not utilize the published astrospheric measurements of
Proxima Cen and 40 Eri A, which gave only upper limits on $\dot{M}$,
and $\lambda$ And and DK UMa, which had uncertain detections of
astrospheric H~I Ly$\alpha$ absorption \citep{Wo05a,Wo05b}.

At the bottom of Table 2 we list three stars that
have parameters at the outer bounds of what we intend to model.
They are test cases for the limits of applicability of the
physical processes summarized in Section \ref{sec:mdot}.
EV Lac is an active M dwarf and flare star that probably has a
fully convective interior \citep[e.g.,][]{Os10}.
Such stars may exhibit qualitatively different mechanisms of mass
loss and rotation-activity correlation than do stars higher up
the main sequence \citep{Mu96,RB07,Ir11,MA11,Vi11}.
V Hya is an N-type carbon star with an extended and asymmetric
AGB envelope and evidence for rapid rotation \citep{Bb95,Kn99}.
89 Her is a post-AGB yellow supergiant with multiple detections of
circumstellar nebular material \citep{SO69,Bj07}.
It is worthwhile to investigate to what extent the
mass loss mechanisms proposed in this paper could be applicable
to these kinds of stars.

Not all stars in Table 2 have precise measurements
for their rotation period.
For 61 Vir and 70 Oph B, we used published estimates of the
rotation period that were obtained from the known correlation
between rotation and chromospheric Ca~II activity \citep{Ba96}.
For essentially all stars more luminous than $\sim 5 \, L_{\odot}$
\citep[with the exception of HR 6902; see][]{Gf88} we estimated
$P_{\rm rot}$ via spectroscopic determinations of $v \sin i$
from the rotational broadening of photospheric absorption lines.
The inclination angle $i$ is the main unknown quantity.
\citet{CM50} found that for an isotropically distributed set of
inclination vectors, the mean value of $\sin i$ is $\pi/4$.
Thus, we estimate a mean rotation period
\begin{equation}
  \langle P_{\rm rot} \rangle \, = \,
  \frac{2\pi R_{\ast}}{(4 / \pi) \, v \sin i}  \,\, .
  \label{eq:Protavg}
\end{equation}
Note that the {\em median} of $\sin i$ for an isotropic distribution
is not equal to the mean; the former is given by $\sqrt{3}/2$.
In order to encompass both values, as well as the majority of
``most likely'' values of $P_{\rm rot}$, we can adopt generous
uncertainty limits for which we will estimate $f_{\ast}$ and the
other derived quantities for mass loss.
For the isotropic distribution of direction vectors,
the quantity $\sin i$ falls between 0.5 and 1 approximately 87\%
of the time.
This is a reasonably good definition for uncertainty bounds that
would correspond to $\pm 1.5 \sigma$ if the distribution were Gaussian.
Thus, for stars with only $v \sin i$ measurements, we use the
following values as error bars on the derived rotation period:
\begin{equation}
  \frac{2}{\pi} \lesssim
  \frac{P_{\rm rot}}{\langle P_{\rm rot} \rangle}
  \lesssim \frac{4}{\pi}  \,\, .
  \label{eq:Protrange}
\end{equation}

\subsection{Comparing Predictions with Observations}
\label{sec:results:compare}

We applied the combined model for mass loss that culminated in
Equation (\ref{eq:combine}) to the stars listed in Table 2.
Below we show results of direct forward modeling; i.e.,
utilizing a known relationship for $f_{\ast}$ as a function of
Rossby number.
First, however, we wanted to investigate whether or not a single
monotonic relationship for $f_{\ast} (\mbox{Ro})$ could produce
mass loss rates that were even remotely close to the measured values.
Thus, we produced trial grids of models in which $f_{\ast}$ was
treated as a free parameter.
For each star, we varied $f_{\ast}$ from $10^{-5}$ to 1 and found
the empirical value of the filling factor ($f_{\rm emp}$) for which
the modeled value of $\dot{M}$ matched the observed value given in
Table 2.
For the four binaries that have only systemic measurements of
$\dot{M}$ we summed the model predictions for each component and
made a single comparison with the observations.

\begin{figure}
\epsscale{1.08}
\plotone{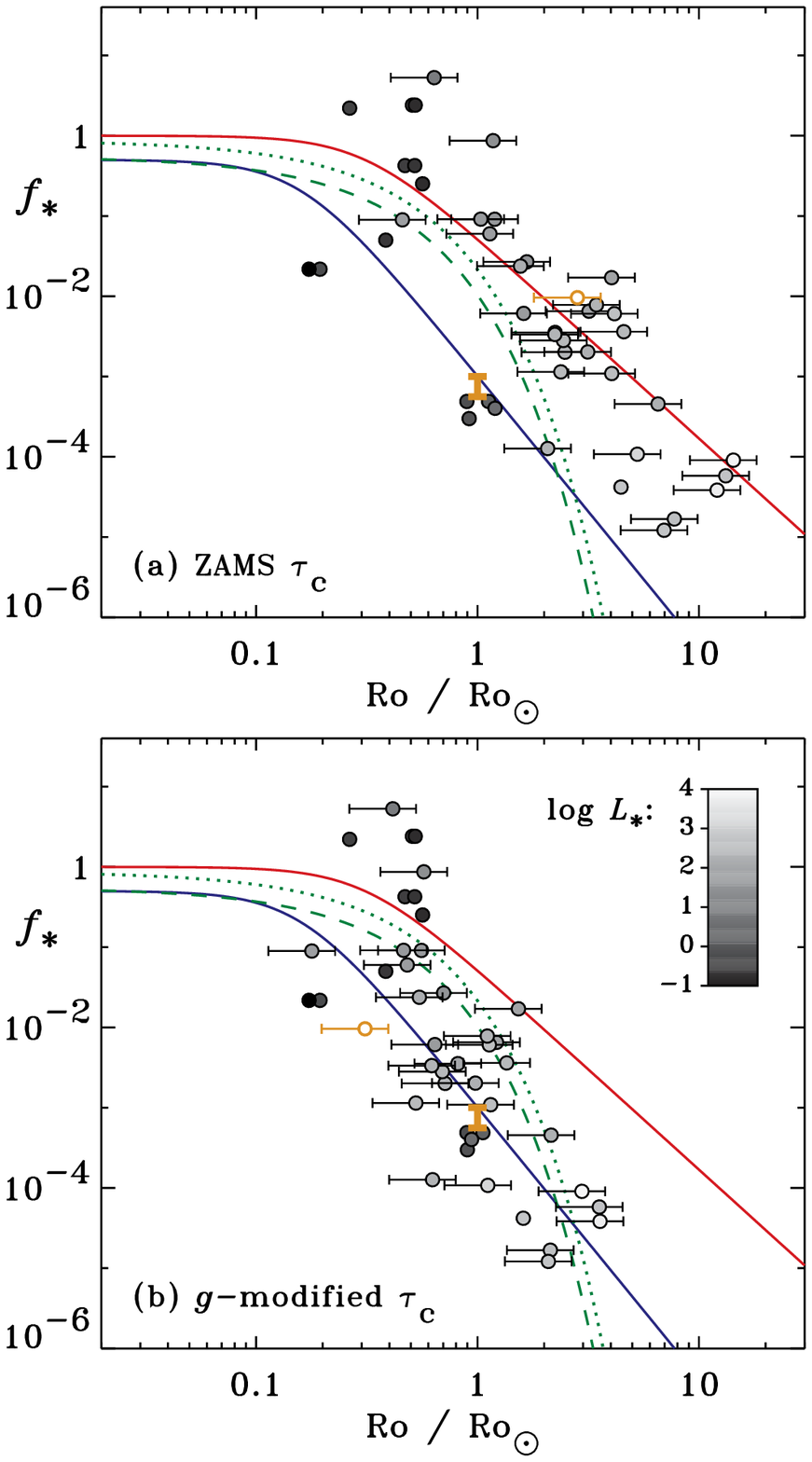}
\caption{Empirical $f_{\rm emp}$ filling factors computed to match
measured mass loss rates.
Rossby numbers were computed in two ways:
(a) directly from Equation (\ref{eq:gunn}), and
(b) multiplying $\tau_c$ from Equation (\ref{eq:gunn}) by
$(g_{\odot}/g)^{0.18}$.
Symbol shading is proportional to $\log (L_{\ast} / L_{\odot})$,
and the curves are the same as those in Figure \ref{fig07}(b).
The thick orange bar denotes the Sun's empirical range of values
(computed from the small variation in $\dot{M}$).
The open orange circle denotes V Hya, the only one of the three
test cases (at bottom of Table 2) that gave a realistic
solution for $f_{\rm emp}$.
\label{fig09}}
\end{figure}

Figure \ref{fig09} shows the result of this process of
``working backwards'' from the measured mass loss rates.
The empirically constrained $f_{\rm emp}$ values are plotted against
Rossby number, which is defined with (a) the simple \citet{Gu98}
ZAMS value for $\tau_c$ (Equation (\ref{eq:gunn})) and (b) a
gravity-modified version of $\tau_c$ that gives giants larger
convective overturn times.\footnote{%
We do not show $f_{\rm emp}$ for the test-case stars EV Lac or
89 Her, since no values in the range $10^{-5}$--1 produced
agreement with their observed mass loss rates.  Extrapolating from
the grid of modeled $\dot{M}$ values to the observed value
would have required impossible values of $f_{\rm emp} \gtrsim 1000$.
We note, however, that the F supergiant 89 Her ``wants'' to
be in the unpopulated upper-right of the $f_{\ast}(\mbox{Ro})$
diagram just like the F6 main sequence star HD 68456 \citep{An10}.
This may be relevant for deducing the relevant physical processes
in other F-type stars with $T_{\rm eff} \approx 6500$ K.}
We varied the exponent in the gravity modification term and found
that multiplying the ZAMS $\tau_c$ by $(g_{\odot}/g)^{0.18}$
gives the narrowest distribution of $f_{\ast}$ versus Ro.
The optimal exponent of 0.18 is very close to the value of 0.23
that we found reproduced the results of \citet{Ha94} and
\citet{Go05,Go07}.
In Figure \ref{fig09} we also show the same curves from
Figure \ref{fig07}(b) that outline the measured range of filling
factors.
The lower envelope curve $f_{\rm min}$, defined in
Equation (\ref{eq:fmin}), appears to be a good match to the
gravity-modified empirical values $f_{\rm emp}$.
This provides circumstantial evidence that $f_{\rm min}$ is indeed
an appropriate proxy for the filling factor of {\em open flux tubes}
as a function of Rossby number.

We now put aside the empirical estimates for the filling factor
and use only Equation (\ref{eq:fmin}) for
$f_{\ast} = f_{\rm min}$ in the remainder of this paper.
Table 3 shows some of the predicted properties of
stellar coronae and winds for the 47 stars in our database.
There were only seven stars for which Equation (\ref{eq:fmin})
gave a filling factor below the adopted ``floor'' value of
$f_{\rm basal} = 10^{-4}$; we replaced $f_{\rm min}$ with
$f_{\rm basal}$ in those cases.
Table 3 also gives $F_{\rm H, TR}$, the coronal heat
flux deposited at the TR for each star.
It may be useful to use this to predict the X-ray flux
associated with open-field regions on these stars, but we
should note that the closed-field regions (which we do not model)
are likely to dominate the observed X-ray emission.
We also list the various components of Equation (\ref{eq:combine})
so that the contributions of gas pressure and wave pressure can
be assessed (see below).
\begin{deluxetable*}{cccccccc}
\tabletypesize{\scriptsize}
\tablenum{3}
\tablecaption{Theoretical Wind Properties of Cool Stars
\label{table03}}
\tablewidth{0pt}

\tablehead{
\colhead{Name} &
\colhead{Ro} &
\colhead{$\log f_{\rm min}$} &
\colhead{$B_{\ast}$ (G)} &
\colhead{$\log F_{H, TR}$} &
\colhead{$\dot{M}_{\rm hot} / \dot{M}_{\rm cold}$} &
\colhead{$\log M_{\rm A, TR}$} &
\colhead{$-\log (\dot{M} / [M_{\odot} \mbox{yr}^{-1}])$}
}

\startdata

Sun                      &
1.960  & -2.996 & 1513.05 & 6.14 &  19.73 & -2.42 & 13.44 \\

$\alpha$ Cen A (G2 V)    &
2.074  & -3.078 & 1308.79 & 6.17 &  11.43 & -2.17 & 13.21 \\

$\alpha$ Cen B (K0 V)    &
1.755  & -2.835 & 1545.97 & 5.62 &  51.73 & -2.93 & 14.09 \\

70 Oph A (K0 V)          &
0.996  & -2.025 & 1666.45 & 5.99 & 194.1  & -3.25 & 13.43 \\

70 Oph B (K5 V)          &
1.027  & -2.067 & 2130.57 & 4.55 & 417.3  & -4.50 & 15.17 \\

$\epsilon$ Eri (K4.5 V)  &
0.519  & -1.174 & 1832.80 & 5.97 & 977.4  & -3.94 & 13.31 \\

61 Cyg A (K5 V)          &
1.107  & -2.173 & 2145.92 & 4.59 & 246.1  & -4.43 & 15.15 \\

$\epsilon$ Ind (K5 V)    &
0.755  & -1.646 & 1941.88 & 5.15 & 551.7  & -4.22 & 14.25 \\

36 Oph A (K5 V)          &
0.923  & -1.918 & 1927.79 & 5.67 & 228.1  & -3.63 & 13.84 \\

36 Oph B (K5 V)          &
1.021  & -2.059 & 1973.54 & 5.51 & 173.1  & -3.66 & 14.16 \\

$\xi$ Boo A (G8 V)       &
0.381  & -0.853 & 1788.75 & 6.69 &  1335  & -3.59 & 12.42 \\

$\xi$ Boo B (K4 V)       &
0.340  & -0.755 & 2620.92 & 4.83 &  5964  & -5.37 & 14.68 \\

61 Vir  (G5 V)           &
1.765  & -2.843 & 1524.08 & 5.99 &  27.96 & -2.63 & 13.56 \\

$\delta$ Eri (K0 IV)     &
1.844  & -2.907 & 1077.96 & 5.77 &  12.04 & -2.18 & 12.72 \\

$\alpha$ Boo (K1.5 III)  &
4.217  & -4.000 &  411.54 & 5.23 &   0.57 & -0.31 & 10.33 \\

$\alpha$ Tau (K5 III)    &
4.183  & -4.000 &  270.35 & 5.17 &   0.75 &  0.11 &  9.88 \\

$\gamma$ Dra (K5 III)    &
4.097  & -4.000 &  301.38 & 5.24 &   0.85 & -0.12 &  9.99 \\

HR 6902 (G9 IIb)         &
3.157  & -3.692 &  287.82 & 6.11 &   1.28 &  0.13 &  9.86 \\

$\beta$ And (M0 III)     &
1.229  & -2.321 &  311.31 & 5.61 &   0.47 & -0.85 &  8.86 \\

$\beta$ UMi (K4 III)     &
6.960  & -4.000 &  262.45 & 5.32 &   0.85 &  0.28 &  9.72 \\

$\mu$ UMa (M0 III)       &
2.182  & -3.152 &  165.83 & 5.81 &   0.63 &  0.68 &  7.93 \\

$\alpha$ TrA (K2 II-III) &
7.010  & -4.000 &  270.87 & 5.55 &   1.25 & -0.13 &  9.32 \\

$\lambda$ Vel (K4 Ib-II) &
5.808  & -4.000 &  136.66 & 5.67 &   2.15 &  1.05 &  8.47 \\

BD $+$01 3070 (RGB)      &
1.122  & -2.192 &  792.36 & 6.49 &   2.36 & -1.48 & 10.14 \\

BD $+$05 3098 (RGB)      &
1.600  & -2.700 &  553.26 & 6.37 &   0.50 & -0.63 &  9.08 \\

BD $+$09 2574 (RGB)      &
3.001  & -3.618 &  632.02 & 5.74 &   0.49 & -0.62 & 10.17 \\

BD $+$09 2870 (RGB)      &
2.249  & -3.196 &  476.21 & 6.00 &   0.37 & -0.17 &  8.68 \\

BD $+$10 2495 (RGB)      &
2.390  & -3.284 &  602.67 & 6.01 &   0.51 & -0.62 &  9.83 \\

BD $+$12 2547 (AGB)      &
2.657  & -3.439 &  378.07 & 5.98 &   0.60 &  0.10 &  9.20 \\

BD $+$17 3248 (RHB)      &
1.259  & -2.355 &  493.93 & 6.81 &   0.65 & -0.53 &  9.05 \\

BD $+$18 2757 (AGB)      &
1.400  & -2.508 &  384.43 & 6.55 &   0.32 & -0.03 &  8.05 \\

BD $+$18 2976 (RGB)      &
2.218  & -3.175 &  464.10 & 5.96 &   0.32 & -0.12 &  8.54 \\

BD $-$03 5215 (RHB)      &
1.095  & -2.157 &  583.84 & 7.06 &   1.91 & -0.90 &  9.36 \\

HD 083212 (RGB)          &
1.361  & -2.467 &  464.86 & 6.20 &   0.23 & -0.44 &  8.07 \\

HD 101063 (SGB)          &
0.813  & -1.744 & 1312.86 & 6.26 &  40.43 & -2.65 & 11.31 \\

HD 107752 (AGB)          &
2.171  & -3.145 &  524.60 & 6.08 &   0.42 & -0.37 &  9.05 \\

HD 110885 (RHB)          &
0.909  & -1.897 &  575.05 & 7.06 &   2.02 & -0.97 &  9.17 \\

HD 111721 (RGB)          &
1.379  & -2.486 &  609.98 & 6.42 &   0.61 & -0.91 &  9.64 \\

HD 115444 (RGB)          &
1.919  & -2.965 &  491.45 & 6.14 &   0.34 & -0.31 &  8.82 \\

HD 119516 (RHB)          &
0.945  & -1.950 &  483.08 & 7.24 &   1.29 & -0.62 &  8.80 \\

HD 121135 (AGB)          &
1.071  & -2.126 &  502.44 & 6.68 &   0.60 & -0.68 &  8.47 \\

HD 122956 (RGB)          &
1.218  & -2.309 &  444.03 & 6.27 &   0.19 & -0.38 &  7.88 \\

HD 135148 (RGB)          &
1.033  & -2.075 &  394.38 & 5.99 &   0.07 & -0.25 &  6.98 \\

HD 195636 (RHB)          &
0.350  & -0.779 &  435.75 & 7.64 &   3.38 & -0.88 &  7.90 \\

\hline

EV Lac (M3.5 V)          &
0.0706 & -0.313 & 3005.25 & 2.72 &   2641 & -8.15 & 17.78 \\

V Hya (N6, AGB)          &
0.609  & -1.367 &   12.39 & 6.53 &   1.32 &  1.99 &  4.34 \\

89 Her (F2 Ib)           &
27.12  & -4.000 &   43.59 & 1.92 &   0.05 & -1.11 & 11.15 \\
\enddata
\end{deluxetable*}

\begin{figure}
\epsscale{1.13}
\plotone{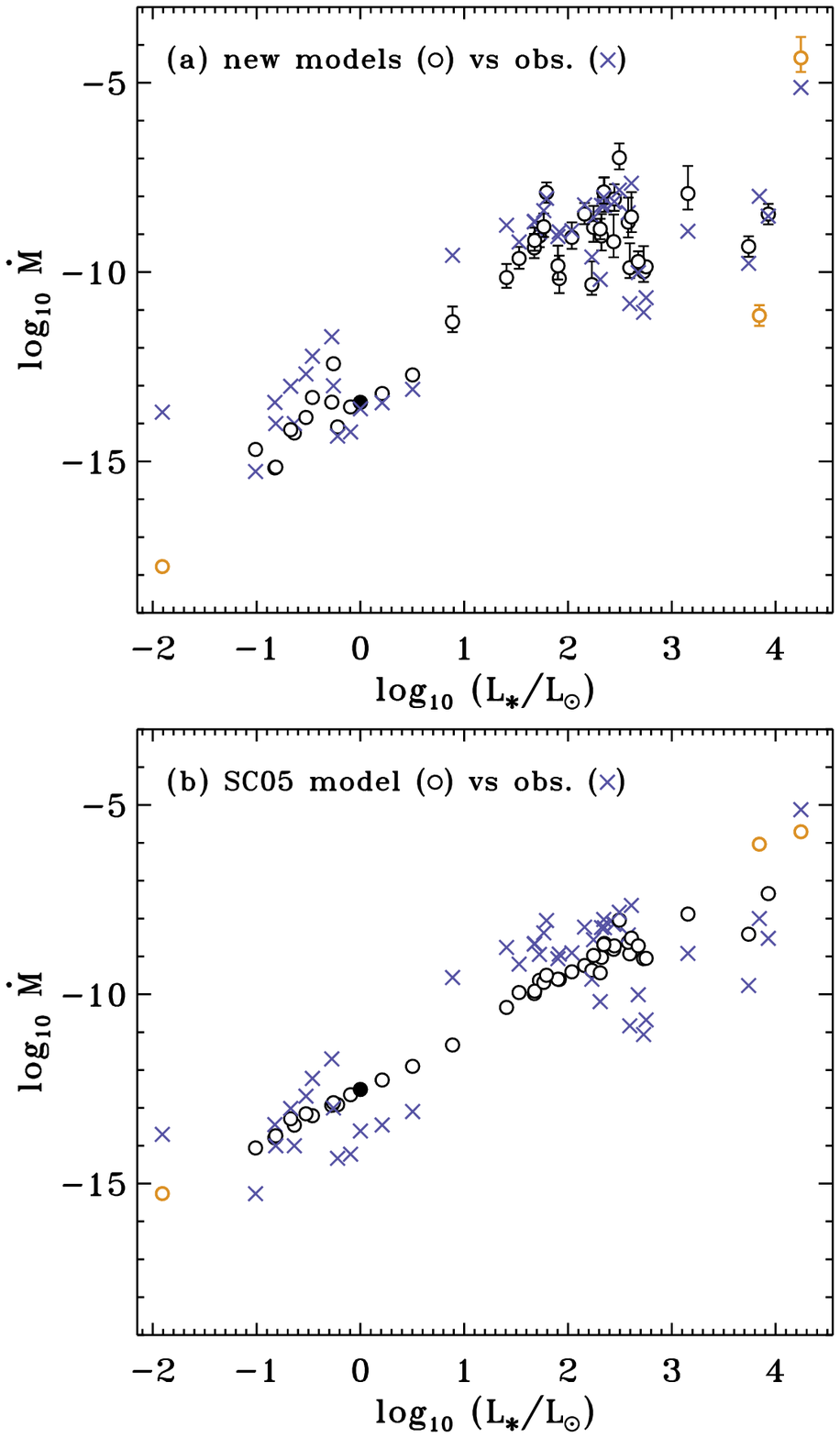}
\caption{Comparison of modeled (open circles) and measured
(blue crosses) mass loss rates for the stars in Table 2,
plotted as $\log \dot{M}$ versus stellar luminosity.
Panels show (a) the standard model developed in this paper,
and (b) the empirical scaling relation of \citet{SC05}.
The Sun is shown as a filled black circle, and the three test-case
stars from the bottom of Table 2 are shown in orange.
Vertical error bars in (a) correspond to models computed for the
$P_{\rm rot}$ range of Equation (\ref{eq:Protrange}).
\label{fig10}}
\end{figure}

Figure \ref{fig10} compares the theoretical and measured mass loss
rates with one another as a function of $L_{\ast}$.
For the four binary systems listed in Table 2 with
combined A$+$B mass loss rates, we separated the measured
value into two pieces using the modeled $\dot{M}$ ratio for the
two components.
(This was done for this figure only because the measured rates
are shown as a function of a single star's luminosity.)
For stars with only $v \sin i$ rotation period estimates, we used
Equation (\ref{eq:Protavg}) to compute $\dot{M}$ for the
central plotting symbol and the entries in Table 3,
and we recomputed $\dot{M}$ for the lower and upper limits given
by Equation (\ref{eq:Protrange}) to obtain the error bars.
Figure \ref{fig10}(b) shows a comparison with the semi-empirical
scaling law proposed by \citet{SC05}, with
\begin{equation}
  \dot{M}_{\rm SC} \, = \, \eta \frac{L_{\ast} R_{\ast}}{M_{\ast}}
  \left( \frac{T_{\rm eff}}{4000 \, \mbox{K}} \right)^{3.5}
  \left( 1 + \frac{g_{\odot}}{4300 \, g} \right)
  \label{eq:SC05}
\end{equation}
where $L_{\ast}$, $R_{\ast}$, and $M_{\ast}$ are assumed to be in
solar units.
For this plot we computed the normalization constant $\eta$ such
that the average modeled mass loss rate would equal the average
measured mass loss rate for all 47 stars.
Averages were taken using the logarithm of $\dot{M}$ so that
all stars would contribute to the average comparably.
We found $\eta = 8.5 \times 10^{-14}$ $M_{\odot}$ yr$^{-1}$, which
is within the error bars of the \citet{SC05} value.

Overall, our ``standard model'' (i.e., Equation (\ref{eq:combine})
with $\alpha_{0} = 0.5$, $h = 0.5$, and $\theta = 1/3$) appears to
match the measured mass loss rates reasonably well.
We emphasize that this model does not contain any arbitrary $\eta$
normalization factors.
For the three test-case stars at the bottom of Table 2,
however, our model does not do as well.
The model underpredicts the mass loss from the dM flare star EV Lac
by at least four orders of magnitude, and it also fails for the
F supergiant 89 Her by a slightly smaller amount.
For EV Lac and other flare-active M dwarfs, it is possible that 
coronal mass ejections and other episodic sources of energy
\citep{Mu96}
could be responsible for the bulk of the observed mass loss.
For the carbon star V Hya, the reasonably good agreement between the
model and measurements is probably a coincidence, since our model
does not include the dusty radiative transfer or strong radial
pulsations that are likely to be important for AGB stars.

It is interesting to highlight the case of the moderately
rotating K dwarfs $\epsilon$ Eri, 70 Oph, and 36 Oph, which
\citet{HJ07} found to have anomalously high mass loss rates.
They concluded that the observed magnetic fluxes for these stars
were insufficient to produce their dense outflows.
For these stars, our modeled mass loss rates tended to be about
a factor of 10--20 below the measured values.
However, these models were computed using $f_{\rm min}$ from
Equation (\ref{eq:fmin}).
The measured values of $f_{\ast}$ given in Table 1
for these stars are larger than their corresponding $f_{\rm min}$
values by factors ranging from 3 to 20.
If instead these values were used, our modeled mass loss rates
would be in better agreement with the astrospheric observations
of \citet{Wo05a}.

We also developed a statistical measure of how well a given model
agrees with the measured database of mass loss rates.
We defined a straightforward least-squares parameter
\begin{equation}
  \chi^{2} \, = \, \frac{1}{N} \sum_{i=1}^{N}
  \left[ \log \dot{M}_{i} (\mbox{model}) -
         \log \dot{M}_{i} (\mbox{obs}) \right]^{2}
\end{equation}
where the total number of comparisons ($N=40$) excludes the final
three test cases in Table 2 and counts each of the four
A$+$B binaries as one.
When $\chi^{2} < 1$, then (on average) the modeled and measured
mass loss rates are within an order of magnitude of one another.
Table 4 summarizes the results, including
comparisons with other published empirical prescriptions.
The $\eta$ normalization factors for each of these scaling laws were
computed similarly as the factor in Equation (\ref{eq:SC05}) above.
Note that our standard model appears to be a significant improvement
over both the popular \citet{Rm75,Rm77} and \citet{SC05} scalings.
\begin{deluxetable}{lc}
\tabletypesize{\small}
\tablenum{4}
\tablecaption{Mass Loss ``Goodness of Fit''
\label{table04}}
\tablewidth{0pt}

\tablehead{
\colhead{Model} &
\colhead{$\chi^{2}$}
}

\startdata

{\bf This paper (standard model)} & {\bf 0.650} \\

This paper (ZAMS $\tau_c$) & 1.575 \\

This paper ($h = 0.25$) & 0.794 \\

This paper ($h = 1$) & 0.564 \\

This paper ($h = 3$) & 0.504 \\

This paper ($\theta = 0.2$) & 0.620 \\

This paper ($\theta = 0.5$) & 0.707 \\

This paper (all [Fe/H]~$=0$) & 0.647 \\

This paper ($\dot{M} = \dot{M}_{\rm hot} + \dot{M}_{\rm cold}$)
  & 0.703 \\

This paper ($\alpha$ from Trampedach \& Stein 2011) & 0.770 \\

\hline

Reimers (1975, 1977) & 1.260 \\

Mullan (1978), Equation (4a) & 3.768 \\

Nieuwenhuijzen \& de Jager (1990) & 2.356 \\

Catelan (2000), Equation (A1) & 1.924 \\

Schr\"{o}der \& Cuntz (2005) & 1.131 \\
\enddata
\end{deluxetable}

To further explore the proposed model, we varied some of the modeling
parameters described in Section \ref{sec:mdot}.
Varying the TR filling factor exponent $\theta$ did not have much
of an effect on $\chi^{2}$.
However, varying the flux height scaling factor $h$ did change
$\chi^{2}$ significantly.
We found that a larger value of $h \approx 3$ gives much better
agreement with the measured mass loss rates than does the standard
value of $h = 0.5$ (see Table 4).
We decided not to adopt this larger value, though, because it falls
well outside the range of empirically determined $h$ values for the
Sun's corona.

We also tried removing some of the imposed complexity of the standard
model to see if simpler assumptions would give adequate results.
Removing the gravity-dependent modification factor of
$(g_{\odot}/g)^{0.18}$ from the definition of the convective overturn
time resulted in significantly poorer agreement with the data
(i.e., more than double the $\chi^2$ of the standard model).
We explored the importance of metallicity by
replacing the published [Fe/H] by purely solar values
([Fe/H]~$= 0$).
This actually improved the value of $\chi^2$ from the standard model,
but only by $< 1$\%.
Removing the exponential factor in Equation (\ref{eq:combine})
gave a slightly higher value of $\chi^2$ (8\% larger than the
standard model).

We also noted that the theoretical photospheric Alfv\'{e}n wave fluxes
from \citet{Mz02a} exhibited a strong dependence on the convective
mixing length parameter (i.e., $F_{\rm A \ast} \propto \alpha^{2.1}$).
Thus, instead of simply assuming $\alpha = 2$ as in the standard model,
we created a linear regression fit to the tabulated simulation results
of \citet{TS11}, who found empirical values of $\alpha$ between 1.6
and 2.2 depending on $T_{\rm eff}$, $\log \, g$, and $M_{\ast}$.
We used the following approximate fit
\begin{equation}
  \alpha_{\rm TS} \, \approx \,
  1.91 - \frac{T_{\rm eff}}{6181 \, \mbox{K}}
  + \frac{\log \, g}{5.58} + \frac{M_{\ast}}{19.1 \, M_{\odot}}
\end{equation}
and did not allow $\alpha_{\rm TS}$ to be less than 1.6 or greater
than 2.2.
Thus, we multiplied the value of $F_{\rm A \ast}$ from
Equation (\ref{eq:FAfit}) by a factor of $(\alpha_{\rm TS}/2)^{2.1}$.
The predicted mass loss rates for the Table 2 stars
had about an 18\% higher value of $\chi^2$ than the standard model,
so we did not pursue this mixing length prescription any further.

\begin{figure}
\epsscale{1.08}
\plotone{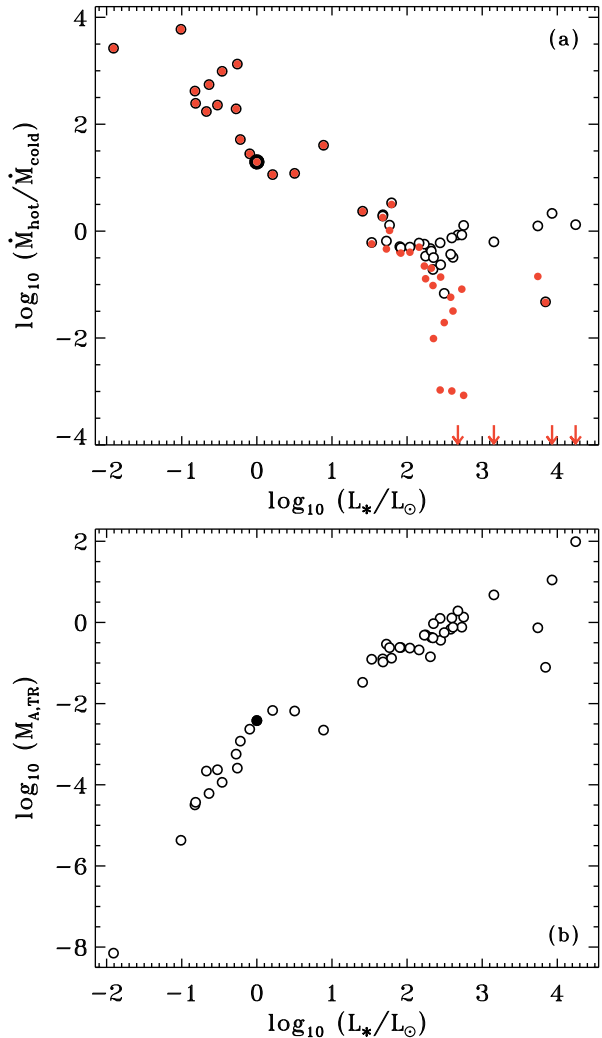}
\caption{Illustrations of the relative importance of ``hot''
versus ``cold'' mass loss mechanisms.
(a) Ratio of hot to cold modeled values for $\dot{M}$ (open circles)
compared with the modified ratio
$\dot{M}_{\rm hot} \exp (-4 M_{\rm A, TR}^{2})/\dot{M}_{\rm cold}$
(red filled circles).
(b) Mach number at the TR, $M_{\rm A, TR} = u_{\rm TR} /
V_{\rm A, TR}$.
The Sun is shown as a thicker black circle in (a) and a filled
black circle in (b).
\label{fig11}}
\end{figure}
For additional context about the hot and cold mass loss models
described in Section \ref{sec:mdot:both}, Figure \ref{fig11}
shows the ratio $\dot{M}_{\rm hot} / \dot{M}_{\rm cold}$ for the
47 modeled stars as well as the TR Mach number
$M_{\rm A, TR} = u_{\rm TR} / V_{\rm A, TR}$.
It is clear that the dwarf stars are dominated by hot coronae,
and the stellar wind outflow is still negligibly small at the
coronal base.
However, as the luminosity exceeds $\sim 50 \, L_{\odot}$ for
the giant stars, the hot coronal contribution goes away and the
acceleration becomes dominated by wave pressure.

\begin{figure}
\epsscale{1.08}
\plotone{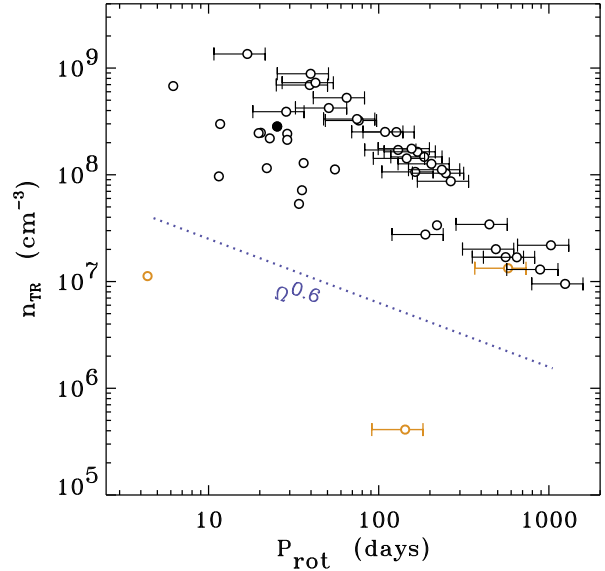}
\caption{Plasma number density at the TR plotted as a function of
stellar rotation period.
Symbols are the same as in Figure \ref{fig10}.
\label{fig12}}
\end{figure}

Figure \ref{fig12} examines how the plasma number density at the
transition region, $n_{\rm TR} = \rho_{\rm TR} / m_{\rm H}$,
depends on stellar rotation.
\citet{HJ07} assumed $n_{\rm TR} \propto \Omega^{0.6}
\propto P_{\rm rot}^{-0.6}$ \citep[see also][]{IT03}.
Although our models do not follow a single universal relation
for both giants and dwarfs, the proposed scaling (or one
slightly steeper) may be appropriate for certain sub-populations
of stars.
For the dwarf stars with well-determined rotation periods, it
is interesting that the Sun's computed value of $n_{\rm TR}$ is
larger than that of stars having higher magnetic activity.
Equation (\ref{eq:rhoTR}) shows that the dependence on filling
factor is weak (i.e., about $f_{\ast}^{0.19}$ for $\theta = 1/3$),
so the variation comes mostly from the other stellar parameters.
This seems to stand in contrast with other observational
determinations of coronal electron densities, where $n$ tends to
increase with activity \citep{Gu04}.
However, X-ray determinations of number density are probably
dominated by closed-field active regions, which are not necessarily
correlated with the regions driving the stellar wind.

\subsection{Predictions for Idealized Stellar Parameters}
\label{sec:results:pre}

In addition to the above comparisons with the individual stars of
Table 2, we also created some purely theoretical sets of
stellar models and computed $\dot{M}$ for them.
We began with the ZAMS model parameters given by \citet{Gi00}, and
we assumed solar metallicity for a range of constant rotation rates.
This gave rise to a two-dimensional grid of models
(varying $T_{\rm eff}$ and $P_{\rm rot}$) for main sequence stars.
Because the modeled stars are all high-gravity dwarfs, we used only
Equation (\ref{eq:gunn}) for $\tau_c$, in combination with
Equation (\ref{eq:fmin}) for $f_{\rm min}$ as a function of
Rossby number.

\begin{figure}
\epsscale{1.13}
\plotone{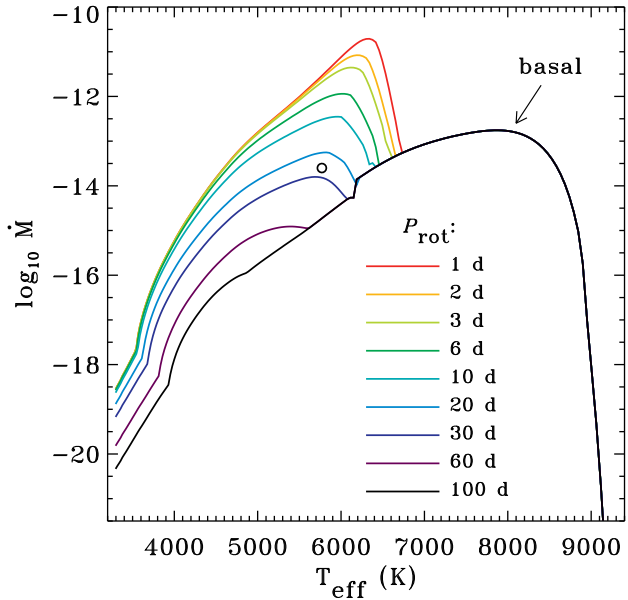}
\caption{Theoretical predictions of $\dot{M}$ for main sequence stars,
plotted as a function of $T_{\rm eff}$.
Differently colored curves show a range of assumed rotation periods
(see labels for values).
The Sun is indicated by an open circle.
\label{fig13}}
\end{figure}

Figure \ref{fig13} shows the resulting mass loss rates as a function
of $T_{\rm eff}$ and $P_{\rm rot}$.
If we had not utilized a basal ``floor'' on $f_{\ast}$, we would
have predicted a steep drop-off in mass loss for
$T_{\rm eff} \gtrsim 7000$~K, at which the \citet{Gu98} expression
for $\tau_{c}$ decreases rapidly.
However, because of the floor, there appear to be reasonably strong
mass loss rates up to the point at which subsurface convection
zones disappear at $T_{\rm eff} \gtrsim 9000$~K.
There is a slightly discontinuous dip in the predicted basal
mass flux around $T_{\rm eff} \approx 6100$~K that arises because
of the iteration for ${\cal R}$, $\rho_{\rm TR}$, and $Q_{\rm TR}$.
If the calculation of these quantities is halted after only one
iteration, the final value of $\dot{M}$ varies more smoothly as a
function of $T_{\rm eff}$.
We plan to utilize a more self-consistent non-WKB model of
Alfv\'{e}n wave reflection in future versions of this work.

The mass loss rates shown in Figure \ref{fig13} are almost all due
to the hot coronal processes discussed in Section \ref{sec:mdot:hot}.
Thus, it is possible to simplify the components of
Equation (\ref{eq:Mdothot}) in order to obtain an approximate
scaling relation for $\dot{M}$ that is reasonable for these main
sequence stellar models.
Ignoring the weakest dependences on some stellar parameters (i.e.,
factors with exponents less than or equal to $1/7$), we found
\begin{displaymath}
  \frac{\dot{M}}{10^{-10} \, M_{\odot}/\mbox{yr}} \, \sim \,
  \left( \frac{R_{\ast}}{R_{\odot}} \right)^{16/7}
  \left( \frac{L_{\ast}}{L_{\odot}} \right)^{-2/7}
\end{displaymath}
\begin{equation}
  \times \,
  \left( \frac{F_{\rm A \ast}}{10^{9} \, \mbox{erg} 
  \,\, \mbox{cm}^{-2} \,\, \mbox{s}^{-1}} \right)^{12/7}
  f_{\ast}^{(4 + 3\theta)/7} \,\, ,
  \label{eq:toosimp}
\end{equation}
which reproduces the curves in Figure \ref{fig13} to within about
an order of magnitude.
Despite the fact that this scaling formula is relatively easy
to apply, we do not recommend its use in stellar evolution or
population synthesis calculations.
Once stars leave the main sequence, Equation (\ref{eq:toosimp})
is no longer a good approximation.

We also computed a time-dependent mass loss rate for the
evolutionary track of a star having $M_{\ast} = 1 \, M_{\odot}$.
There is evidence that the wind from the ``young Sun'' was
significantly denser than it is today, and this more energetic
outflow may have been important to early planetary evolution
\citep[e.g.,][]{Wo06,Gu07,St11,Su11}.
We used the BaSTI evolutionary track plotted in Figure \ref{fig01}
\citep{Pi04} for the time variation of $R_{\ast}$ and $L_{\ast}$.
We grafted on a model of rotational evolution for a solar-mass
star from Figure 6(a) of \citet{Dn10}.
For late ages ($t \gtrsim 100$ Myr, or $\log t \gtrsim 8$), this
model has approximately $P_{\rm rot} \propto t^{0.54}$.
Such an age scaling is well within the range of empirically
determined power laws ($t^{0.5}$ to $t^{0.6}$) obtained
from young solar analogs \citep[e.g.,][]{Ba03,Gu07}.

\begin{figure}
\epsscale{1.12}
\plotone{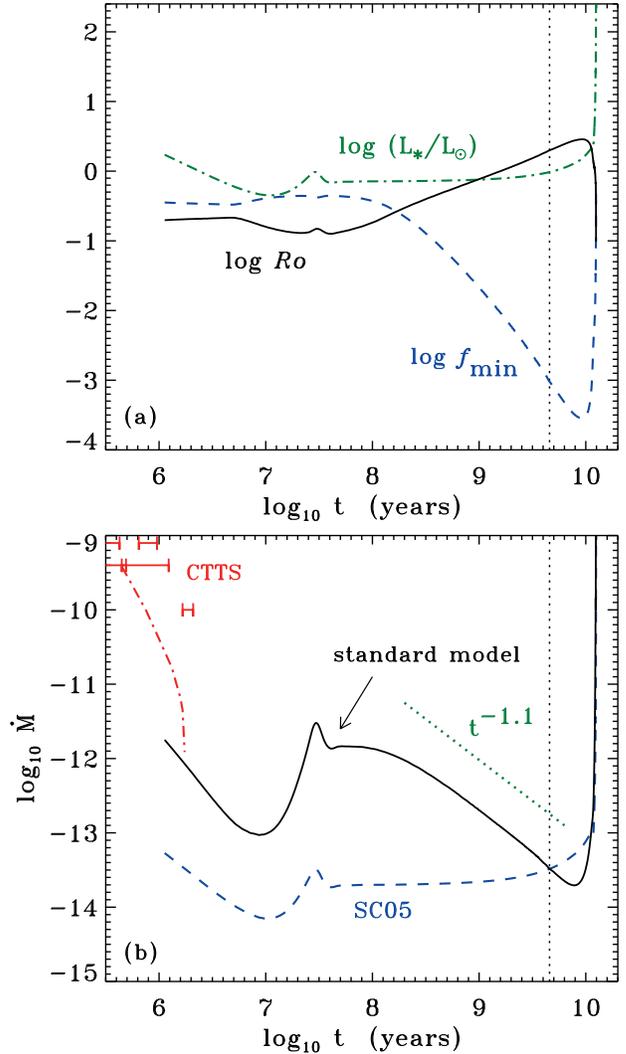}
\caption{Theoretical predictions of stellar wind properties for an
evolving solar-mass star.
(a) Base-10 logarithms of luminosity (green dot-dashed curve),
Rossby number (black solid curve), and $f_{\rm min}$ (blue
dashed curve) plotted as a function of age in years.
(b) Our standard model for $\dot{M}$ (black solid curve), compared
with the \citet{SC05} scaling (blue dashed curve) and an ideal
power-law $t^{-1.1}$ decline with increasing age (green dotted curve).
A model and observations of CTTS are shown for comparison (red
error bars and dot-dashed curve).
\label{fig14}}
\end{figure}

Figure \ref{fig14}(a) shows how the luminosity and two dimensionless
parameters related to the rotational dynamo (Ro and $f_{\rm min}$)
vary as a function of age for this model.
Note that prior to about $t \approx 70$ Myr the Rossby number is
small enough that the filling factor appears to be saturated
near its maximum assumed value of 0.5.
At very late times, when the star begins to ascend the red giant
branch, the Rossby number decreases again because of the
increase in $\tau_c$ with decreasing $T_{\rm eff}$ and gravity.
We utilized the $(g_{\odot}/g)^{0.18}$ correction factor when
computing $\tau_c$, but it was relatively unimportant until the
star left the main sequence.

Figure \ref{fig14}(b) gives our prediction for the age variation
of the a solar-type star's mass loss rate.
For ages between about $t \approx 0.2$ and 7 Gyr the decrease
in mass loss appears to be fit approximately by a power law,
with $\dot{M} \propto t^{-1.1}$.
This is a significantly shallower age dependence than the
$t^{-2}$ decline suggested by \citet{Wo02} on the basis of
astrosphere measurements.
We note that if the rotation period was the only variable to
change with time, Equation (\ref{eq:toosimp}) would give
something like $\dot{M} \propto P_{\rm rot}^{-2.4}$
(for $f_{\ast} = f_{\rm min}$ and $\theta = 1/3$).
Thus, a more rapid increase of $P_{\rm rot}$ with age---such as
the $t^{0.85}$ dependence in the solar-mass rotational model
of \citet{Ln10}---would give rise to a steeper age-$\dot{M}$
relationship more similar to that of \citet{Wo02}.

For comparison, Figure \ref{fig14}(b) also shows that the
\citet{SC05} scaling law predicts a much smaller range of mass
loss variation for the young Sun than does the present model.
We also show a model \citep{Cr08,Cr09} and measurements \citep{HEG}
for classical T Tauri stars (CTTS) at the youngest ages.
It is clear that for $t \lesssim 10$ Myr some additional physical
processes must be included (e.g., accretion-driven turbulence
on the stellar surface) to successfully predict mass loss rates.

\section{Discussion and Conclusions}
\label{sec:conc}

The primary aim of this paper was to develop a new generation of
physically motivated models of the winds of cool
main sequence stars and evolved giants.
These models follow the production of MHD turbulent motions from
subsurface convection zones to their eventual dissipation and
escape through the stellar wind.
The magnetic activity of these stars is taken into account by
extending standard age-activity-rotation indicators to include
the evolution of the filling factor of strong magnetic fields in
stellar photospheres.
The winds of G and K dwarf stars tend to be driven by gas pressure
from hot coronae, whereas the cooler outflows of red giants are
supported mainly by Alfv\'{e}n wave pressure.
We tested our model of combined ``hot'' and ``cold'' winds by
comparing with the observed mass loss rates of 47 stars, and we
found that this model produces better agreement with the data than
do published scaling laws.
We also made predictions for the parametric dependence of
$\dot{M}$ on $T_{\rm eff}$ and rotation period for main sequence
stars, and on age for a one solar mass evolutionary track.

The eventual goal of this project is to provide a straightforward
algorithm for predicting the mass loss rates of cool stars for use
in calculations of stellar evolution and population synthesis.
A brief stand-alone subroutine called BOREAS has been developed to
implement the model described in this paper.
This code is written in the Interactive Data Language (IDL)\footnote{%
IDL is published by ITT Visual Information Solutions.  There are
also several free implementations with compatible syntax, including
the GNU Data Language (GDL) and the Perl Data Language (PDL).}
and it is included with this paper as online-only material.
This code is also provided, with updates as needed, on the first
author's web page.\footnote{http://www.cfa.harvard.edu/$\sim$scranmer/}
Packaged with the code itself are data files that allow the user to
reproduce many of the results shown in Section \ref{sec:results}.

In order to further test the conjecture that Alfv\'{e}n waves and
turbulence drive cool-star winds, the models need to be expanded
from the simple scaling laws of Section \ref{sec:mdot} to fully
self-consistent solutions of the mass, momentum, and energy
conservation equations along open flux tubes.
Modeling the full radial dependence of density, temperature,
magnetic field strength, and outflow speed would eliminate our
reliance on approximate factors like $h$ and $\theta$.
We believe that the models of \citet{CvB07} for the solar wind,
and \citet{Cr08} for T Tauri stars, can be extended
straightforwardly and applied to other types of stars.
However, there are many other approaches to producing self-consistent
and/or three-dimensional models that should be explored
\citep[e.g.,][]{Ai00,Ai10,HJ05,ST05,FG06,Su07,Vi09,Co09,Co11}.

There are additional ways that our simplified models of coronal
energy balance (Section \ref{sec:mdot:hot}) and wave-pressure
driving (Section \ref{sec:mdot:cold}) may be improved:
\begin{enumerate}
\item
Our standard assumption for the outflow speed in a coronal wind
was $u_{\infty} = V_{\rm esc}$.
However, \citet{Ju92} found that many stars have significantly
smaller terminal speeds.
It should be possible to use something like the \citet{SM03}
solar wind scaling law to estimate the peak temperature
in the corona, and thus apply the \citet{P58} theory of gas
pressure acceleration to compute the wind speed.
\item
We assumed in Section \ref{sec:mdot:hot} that
$r_{\rm TR} \approx R_{\ast}$.
However, \citet{SC05} estimated that some low-gravity stars should
exhibit ``puffed up'' chromospheres with a fractional extent given
by the final term in parentheses in Equation (\ref{eq:SC05}).
We applied this correction factor to the modeled values of
$r_{\rm TR}$ and $\dot{M}_{\rm hot}$ for the stars in
Table 2.
Doing so yielded significant differences from the standard model
only for stars having $\dot{M}_{\rm cold} \gg \dot{M}_{\rm hot}$,
i.e., the combined model value of $\dot{M}$ was relatively
unchanged in those cases.
However, in general there may be other stars for which this kind of
correction factor needs to be considered in more detail.
\item
We also assumed that the flux height scaling factor $h$ took on
a single constant value for all stars.
It may be useful to explore extending the \citet{SC05}
idea of gravity-dependent spatial expansion to this parameter
as well.
The $\chi^2$ results shown in Table 4 suggest that
a larger value of $h \approx 3$ could be appropriate for many of
the low-gravity stars in our observational database, whereas
the range $h \approx 0.5$--1 is probably best for main sequence
stars like the Sun.
\item
The flux of energy $F_{\rm A \ast}$ in kink/Alfv\'{e}n waves
in the photosphere may depend on other parameters that we have
not considered.
\citet{Mz02a} found that the flux is rather sensitive to
$B_{\ast}/B_{\rm eq}$ in the photosphere, so departures from our
assumed value of 1.13 may give rise to significantly different
predictions.
Also, \citet{Mz02b} examined the sensitivity to metallicity and
found that lower $Z/Z_{\odot}$ tends to give lower values of
$F_{\rm A \ast}$ for $T_{\rm eff} \lesssim 6000$ K.
Preliminary tests showed that this effect does not strongly
affect the mass loss rates derived in this paper.
However, a varying metallicity should also change other properties
of the convection---including the effective mixing length $\alpha$
parameter---thus possibly making this effect more important.
\item
Instead of assuming a simple monotonic dependence of the open-field
filling factor on Rossby number, it may be possible to construct
realistic surface distributions of active regions for a given
activity level and rotation period, and model the opening up of
flux tubes by both stellar winds and centrifugal forces
\citep{Mu83,MS87,Ja04,HJ05,Co09}.
\end{enumerate}
To continue testing and refining these models, it is also important
to utilize the newest and most accurate measurements of stellar
mass loss rates \citep[see, e.g.,][]{SC07,Wi09,Ca09,MJ11,Vy11} and
magnetic fields \citep{DL09,Vl11}.

Finally, we emphasize that a complete description of late-type
stellar winds requires the incorporation of other physical
processes besides Alfv\'{e}n waves and turbulence.
The outer atmospheres of cool stars are also likely to be powered
by acoustic or longitudinal MHD waves \citep{Cu90,Bu98},
episodic flares or coronal mass ejections \citep{Mu96,Aa09},
and large-amplitude pulsations \citep{Bw88,dJ97,Wi00}.
It is well known that radiative driving should not be neglected
for AGB stars and red supergiants, and it may be important for
Cepheids \citep{NL08} and horizontal branch stars \citep{VC02}
as well.

\acknowledgments

The authors gratefully acknowledge Nancy Brickhouse,
Andrea Dupree, Adriaan van Ballegooijen, Stan Owocki, Ofer Cohen,
and the anonymous referee for many valuable discussions.
This work was supported by the Sprague Fund of the Smithsonian
Institution Research Endowment, and by the National Aeronautics
and Space Administration (NASA) under grants {NNX\-09\-AB27G} and
{NNX\-10\-AC11G} to the Smithsonian Astrophysical Observatory.
This research made extensive use of NASA's Astrophysics
Data System and the SIMBAD database operated at CDS,
Strasbourg, France.
This research has also made use of the NASA/IPAC/NExScI Star and
Exoplanet Database (NStED), which is operated by the Jet Propulsion
Laboratory, California Institute of Technology, under contract with
NASA.

\appendix
\section{Notes on Stellar Magnetic Field Measurements}
\label{appen:bf}

In this work we focus on observations of unpolarized spectral lines
sensitive to Zeeman splitting by stellar magnetic fields
\citep[e.g.,][]{Rb80}.
The resulting ``Zeeman broadened'' line profiles are valuable probes
of both the intensity-weighted mean absolute value of the field
strength (i.e., $B_{\ast} = \langle I_{\rm B} |B| \rangle /
\langle I_{\rm B} \rangle$, where $I_{\rm B}$ is the continuum
intensity in the magnetic regions)
and the intensity-weighted fraction of the visible stellar hemisphere
that is covered by these fields (i.e., the filling factor $f_{\ast}$).
These detections are thus weighted towards the brightest regions
of the stellar surface; i.e., plage or network regions.
However, this technique allows detection of more topologically complex
fields, and thus more comprehensive values of $f_{\ast}$ and $B_{\ast}$,
than does the use of circular polarization.
The latter exhibits significant signal cancellation when there are
multiple oppositely directed patches of magnetic field in the same
resolution element.  
In many cases, however, only the disk-averaged magnetic flux
density ($B_{\ast} f_{\ast}$) can be determined reliably from
Zeeman broadened spectra and not the separate values of $B_{\ast}$
and $f_{\ast}$ \citep[see also][]{Ru97,An10}.

Many details of the observations of the stars discussed in
Section \ref{sec:mag} were given by \citet{SL85,SL86} and
\citet{Sa90,Sa91,Sa96a,Sa96b,Sa01}.
The approximate quality factors listed in Table 1
span the range from low ($q=1$) to high ($q=4$) relative confidence
in the derived magnetic parameters.
The values for $q$ were assigned based on a combination of the
following properties of the spectroscopic data and its magnetic
analysis:
\begin{enumerate}
\item
Detections using lines with longer wavelengths and higher
Land\'{e} $g_{\rm eff}$ factors are given higher $q$ values.
The ratio of the strength of the Zeeman effect to the nonmagnetic
Doppler width is $\Delta \lambda_{\rm B} / \Delta \lambda_{\rm D}$,
where $\Delta \lambda_{\rm B} \propto g_{\rm eff} \lambda^2$ is the
Zeeman splitting amplitude and
$\Delta \lambda_{\rm D} \propto \lambda$ is the nonmagnetic Doppler
width.
Thus, the relative detectability of the effect increases as
$g_{\rm eff} \lambda$.
\item
Spectra with higher signal-to-noise (S/N) ratios and higher
spectral resolution (${\cal R} = \lambda / \Delta \lambda$) tend
to have higher quality \citep[see, e.g.,][]{Sa88}, although
a longer wavelength measurement can trump better ${\cal R}$.
For example, the greater magnetic sensitivity at large $\lambda$
can lead to partial resolution of the individual Zeeman components
\citep{SL85}. 
\item
The simultaneous analysis of larger numbers of lines (especially
with higher $g_{\rm eff}$) contributes to a good quality measurement
\citep{Ru97}.
It is additionally helpful for the lines to be free of blends and
for any rotational broadening to be small (i.e.,
$v \sin i \lesssim 10$ km s$^{-1}$); see also \citet{Sa88}.
\item
Finally, the method and quality of line modeling---including the
level of detail in the magnetic radiative transfer, the atmospheric
models used, details of integration over the stellar disk, and the
treatment of line blends---vary widely.
This field has seen a continual improvement in the sophistication
of these models, with an attendant increase in our understanding
of the measurement and systematic uncertainties
\citep[e.g.,][]{An10,Sy10}.
\end{enumerate}
We chose to exclude several published observations from the list of
stars given in Table 1.
We did not use the M dwarf data presented by \citet{Rn09} because
these stars all tended to sit in the saturated region of
Figure \ref{fig07} (i.e., $\mbox{Ro}/\mbox{Ro}_{\odot} < 0.1$),
and thus do not contribute to improving our knowledge of the
rotation dependence of $f_{\ast}$.
We did not include the measured field of the F6 main sequence star
HD 68456 \citep{An10} because we do not attempt to model the
outflows of stars significantly hotter than the Sun.
Also, its combined strong field and high Rossby number point to
a possible transition to a different type of magnetic activity from
that described by the standard cool-star age-activity-rotation
relationship \citep[see also][]{BV02}.

\end{document}